\documentclass{aastex62}
\usepackage[utf8]{inputenc}
\usepackage{gensymb}

\graphicspath{{./}{figures/}}

\submitjournal{ApJ}
\received{December 31, 2020}
\revised{January 26, 2021}
\accepted{February 1, 2021}

\shorttitle{Dynamical coupling of mean-field dynamo and wind}
\shortauthors{Perri et al.}

\begin{document}

\title{Dynamical coupling of a mean-field dynamo and its wind: Feedback loop over a stellar activity cycle}

\correspondingauthor{Barbara Perri}
\email{barbara.perri@kuleuven.be}

\author[0000-0002-2137-2896]{Barbara Perri}
\affil{Université Paris Saclay and Université de Paris, CEA, CNRS, \textbf{AIM}, F-91190 Gif-sur-Yvette, France}
\affil{Centre for mathematical Plasma Astrophysics, KU Leuven, Celestijnenlaan 200b-box 2400, 3001 Leuven, Belgium}

\author[0000-0002-1729-8267]{Allan Sacha Brun}
\affiliation{Université Paris Saclay and Université de Paris, CEA, CNRS, \textbf{AIM}, F-91190 Gif-sur-Yvette, France}

\author[0000-0002-9630-6463]{Antoine Strugarek}
\affiliation{Université Paris Saclay and Université de Paris, CEA, CNRS, \textbf{AIM}, F-91190 Gif-sur-Yvette, France}

\author[0000-0002-2916-3837]{Victor Réville}
\affil{IRAP, Université Toulouse III - Paul Sabatier, CNRS, CNES, Toulouse, France}

\begin{abstract}
We focus on the connection between the internal dynamo magnetic field and the stellar wind. If the star has a cyclic dynamo, the modulations of the magnetic field can affect the wind which in turn can back-react on the boundary conditions of the star, creating a feedback loop. We have developed a 2.5-dimensional numerical set-up to model this essential coupling. We have implemented an alpha-Omega mean-field dynamo in the PLUTO code and then coupled it to a spherical polytropic wind model via an interface composed of four grid layers with dedicated boundary conditions. We present here a dynamo model close to a young Sun with cyclic magnetic activity. First we show how this model allows to track the influence of the dynamo activity on the corona by displaying the correlation between the activity cycle, the coronal structure and the time evolution of integrated quantities. Then we add the feedback of the wind on the dynamo and discuss the changes observed in the dynamo symmetry and the wind variations. We explain these changes in terms of dynamo modes: in this parameter regime, the feedback loop leads to a coupling between the dynamo families via a preferred growth of the quadrupolar mode. We also study our interface in terms of magnetic helicity and show that it leads to a small injection in the dynamo. This model confirms the importance of coupling physically internal and external stellar layers, as it has a direct impact on both the dynamo and the wind.
\end{abstract}


\section{Introduction}
\label{sec:intro}

Stars are complex objects that require to combine many fields of physics to be understood as a whole. In particular the stellar interiors have traditionally been associated with fluid dynamics and later on magnetohydrodynamics (MHD), while stellar winds and atmospheres are usually described using plasma physics. To go beyond this first approach, these various communities had to gather their knowledge to produce the most consistent models possible. The best example for this is of course the closest star from us and hence the easiest one to observe in thorough details, the Sun.

For more than four centuries the Sun has been observed displaying a cyclic activity through sunspot observations. Later on, \cite{hale_zeeman_1908} linked the sunspots to magnetic activity, which led to realize it has an 11-year period for amplitude and 22-year period for polarity. The general framework adopted nowadays to explain such a generation of large-scale magnetic field is the interface dynamo \citep{moffatt_magnetic_1978, parker_solar_1993, browning_dynamo_2006}: the differential rotation profile in the convection zone of the star \citep{schou_flows_1998, thompson_internal_2003} leads to the generation of strong toroidal fields at the tachochline \citep{spiegel_solar_1992}, which in turn is used to regenerate poloidal fields thanks to the combination of turbulence, buoyancy and Coriolis force at the surface. This dynamo loop allows for the amplification of the initial magnetic field until saturation, thus sustaining it against ohmic dissipation \citep{miesch_large-scale_2005, brun_magnetism_2017}. To analyse the dynamo field, it can be decomposed into modes characterized by degrees $\ell$ and $m$ by projecting it on the spherical harmonics base. Under the assumption of axisymmetry (e.g $m=0$), odd $\ell$ are called dipolar modes and constitutes the primary dynamo family, while even $\ell$ are called quadrupolar modes and constitutes the secondary family \citep{mcfadden_reversals_1991}. A simplified yet efficient approach to model solar magnetism is the mean-field dynamo \citep{roberts_kinematic_1972, krause_mean-field_1980}, focusing on large-scale fields and assuming axisymmetry. The generation of toroidal field through differential rotation is then deemed the $\Omega$ effect and the regeneration of the poloidal or toroidal field via turbulence is deemed the $\alpha$ effect. The combination of these two effects can give $\alpha$-$\Omega$, $\alpha^2$ or $\alpha^2$-$\Omega$ dynamo loops. This description has the advantages of being easily implemented in MHD simulations with low computational costs and yielding realistic results \citep{charbonneau_dynamo_2020, tobias_turbulent_2019}.

On the other hand, it was not until the observations of the \textit{Mariner 2} spacecraft that the existence of a dynamic solar atmosphere being accelerated into a transsonic and transsalfvénic wind was accepted by the community \citep{neugebauer_solar_1962}. The first hydrodynamical description was given by \cite{parker_dynamics_1958} and magnetism was added by \cite{schatzman_theory_1962}, \cite{weber_angular_1967}, \cite{mestel_magnetic_1968} and \cite{sakurai_magnetic_1985} to yield a better description of the corresponding torque applied to the star. The solar wind can be described using empirical models such as the WSA model \citep{wang_magnetic_1990, arge_improvement_2000} linking the terminal speed of a flux tube with its expansion factor. Another way is to use MHD simulations in 1D along a flux tube \citep{lionello_including_2001, suzuki_saturation_2013, pinto_multiple_2017}, in 2D with axisymmetry \citep{keppens_numerical_1999, matt_accretion-powered_2008, reville_solar_2015} or in 3D with a full description of the corona \citep{toth_adaptive_2012, usmanov_three-fluid_2014, riley_inferring_2015, reville_global_2017}. The wind can even be better described using a multi-fluid approach \citep{usmanov_global_2000, hollweg_generation_2002} to take into account the influence of the different populations of particles. There are still open problems like the precise mechanism behind the heating of the corona which can be approximated by a polytropic state with a low adiabatic index, but is described better with detailed heating mechanisms such as magnetic reconnection \citep{parker_nanoflares_1988} or the turbulent dissipation of waves \citep{cranmer_self-consistent_2007, reville_parametric_2018, shoda_alfven-wave_2020} in a non-trivial way. 

Recent observations, by the satellite \textit{Ulysses} for cycle 22 and the beginning of cycle 23 \citep{mccomas_three-dimensional_2003, mccomas_weaker_2008} or by the satellite \textit{OMNI} for cycles 23 and 24 \citep{owens_global_2017}, have shown that there is a correlation between the 11-year dynamo cycle and the evolution of the corona. During a minimum of activity, the magnetic field is low in amplitude and its topology is mostly dipolar. The corona is also very structured with fast wind at the poles (around 800 km/s) associated with coronal holes, and slow wind at the equator (around 400 km/s) associated with streamers \citep{wang_coronal_2007}. During a maximum of activity, the magnetic field is high in amplitude and its topology is a mixture of high modes, dominated by the quadrupolar modes with equatorial symmetry \citep{derosa_solar_2012}, while fast and slow solar streams can be found at all co-latitudes. Complementary observations from Earth using IPS (Inter-Planetary Scintillations, see \cite{hewish_interplanetary_1964} and \cite{asai_heliospheric_1998}) can help reconstruct the latitudinal evolution of the solar wind speed over time. They display an 11-year pattern in the form of a cross due to the correlation between the activity cycle and the latitudinal distribution of slow and fast streams \citep{tokumaru_solar_2010,sokol_reconstruction_2015}. This suggests that there is a coupling operating between the interior and the exterior of the Sun, but the exact physical mechanisms at play as well as the relevant timescales are still unclear \citep{wedemeyer-bohm_coupling_2009}.

From a theoretical and numerical point of view, it is however very difficult to study all of these layers simultaneously. The magnetic field and the wind evolve over a broad range of scales: the magnetic field evolves over 11 year with various scales ranging from several Megameters to a solar radius \citep{brun_magnetism_2017}, while the solar wind adapts in a few hours, accelerating over a couple of tens of solar radii but extending up to several Astronomical Units (AUs) \citep{meyer-vernet_basics_2007}. The physical properties of their respective environment are also very different: the interior of the star is characterized with convection \citep{rieutord_suns_2010} and large scale flows \citep{thompson_internal_2003, basu_characteristics_2010}, while the exterior witnesses eruptive events in a dynamical atmosphere \citep{schrijver_magnetic_2005}. The rapidly changing $\beta$ plasma parameter (ratio of the thermal pressure over the magnetic pressure) at the surface of the star, or the Mach number going from subsonic to supersonic in the corona are also evidence that the relative importance of physical phenomena can vary between the interior of the star and its extended atmosphere \citep{gary_plasma_2001}. 
All of these disparities make the modeling of this coupling a numerical challenge.

There have been many attempts to resolve this problem with various approaches. A first approach is to use a quasi-static approach, meaning that the coupling is realized through a series of wind relaxed states corresponding to a sequence of magnetic field configurations evolving in time. These models can be data-driven, using series of magnetic field observations \citep{luhmann_solar_2002, merkin_time-dependent_2016, reville_global_2017}, or rely on the numerical coupling between two codes dedicated respectively to the inner and outer layers of the Sun \citep{pinto_coupling_2011, perri_simulations_2018}. Another approach is to focus on the surface with a numerical box of a few tens of Megameters to capture the small time and spatial scales. This means that this approach can include all the relevant physical processes (for example convection or radiative transfer at the surface) but on a short period of time and only for a specific region of the Sun \citep{skartlien_excitation_2000, vogler_simulations_2005, stein_solar_2006, wedemeyer-bohm_coupling_2009}, so they lack global coupling. Finally there have been some attempts to model a dynamical coupling on a global scale, for example in \cite{von_rekowski_stellar_2006} or \cite{warnecke_influence_2016}.
The first study focused on T-Tauri-like stars with an accretion disk and showed that the presence of a cyclic magnetic field in the star greatly affects its wind and its interaction with the magnetized disk. The second study focused on the influence of a realistic transition region on solar-like convective zones and showed changes in the differential rotation and the magnetic field evolution, which means that the modeling of the stellar surface can be crucial for its dynamics.

We wish to follow a similar approach of full dynamical coupling of the inner and outer layers of a star with an emphasis on improving the treatment of the physical interface between these two layers. Our aim is to focus on the large-scale field by using mean-field and axisymmetry assumptions. We design a 2.5D numerical model including the star and its corona, and design an interface to control the interaction between the two zones. For this first study we will focus on young solar-like stars to reduce the discrepancies of the time scales between the dynamo and the wind. In section \ref{sec:set_up}, we describe the choices we made for the modeling of the dynamo and the wind, and finally the interface between the two computational domains. 
In section \ref{sec:evol}, we present an application of our model to a case of a young Sun with a short dynamo period. First we show only the influence of the dynamo-generated magnetic field on the structure of the corona and wind integrated quantities such as the Alfvén radius and the mass loss.
Then in section \ref{sec:feedback_loop} we allow the wind to back-react on the dynamo in a two-way coupling, and quantify its impact. We discuss the feedback loop in our simulations by analyzing it in terms of dynamo modes and helicity generation. Finally in section \ref{sec:conclusion} we summarize our main results and discuss the next steps for future studies.

\section{Numerical dynamo-wind set-up}
\label{sec:set_up}

We begin by presenting the equations solved and our numerical set-up dedicated to the study of dynamo-wind coupling. We use the PLUTO code \citep{mignone_pluto:_2007}, which is a compressible multi-physics and multi-solver code with an extended MHD module. We first present the wind model in section \ref{subsec:model_wind}, already described in \cite{perri_simulations_2018}. Then we present the implementation of the mean-field dynamo and its validation in section \ref{subsec:model_dyn}. Finally we describe the numerical interface between the two computational zones in section \ref{subsec:model_interface}. 
We will limit ourselves here to an axisymmetric (2.5D) description of the dynamo-wind coupling. The main consequence is that any magnetic flux emerging through the stellar boundary is axisymmetric, and we cannot cope with magnetic structures localised in longitude such as starspots and active regions. This limitation does not prevent us to assess how the coupling operates on a large-scale basis, and we leave the extension to fully 3D mean-field models \citep{karak_solar_2017, kumar_3d_2019} for future work.

\subsection{Wind model}
\label{subsec:model_wind}

Our wind model is adapted from \cite{reville_solar_2015}. We solve the set of the conservative ideal MHD equations composed of the continuity equation for the density $\rho$, the momentum equation for the velocity field $\mathbf{u}$ with its momentum written $\mathbf{m}=\rho\mathbf{u}$, the energy equation with the energy noted $E$ and the induction equation for the magnetic field $\mathbf{B}$:
\begin{equation}
\frac{\partial}{\partial t}\rho+\nabla\cdot\rho\mathbf{u}=0,
\label{eq:rho_wind}
\end{equation} 
\begin{equation}
\frac{\partial}{\partial t}\mathbf{m}+\nabla\cdot(\mathbf{mu}-\frac{\mathbf{BB}}{4\pi}+\mathbf{I}p) = \rho\mathbf{a},
\label{eq:m_wind}
\end{equation}
\begin{equation}
\frac{\partial}{\partial t}E + \nabla\cdot((E+p)\mathbf{u}-\frac{\mathbf{B}}{4\pi}(\mathbf{u}\cdot\mathbf{B})) = \mathbf{m}\cdot\mathbf{a} -\frac{1}{4\pi} \nabla\cdot\left[\left(\eta_w \left(\nabla\times\mathbf{B}\right)\times\mathbf{B}\right)\right],
\label{eq:E_wind}
\end{equation}
\begin{equation}
\frac{\partial}{\partial t}\mathbf{B}+\nabla\cdot(\mathbf{uB}-\mathbf{Bu})= -\nabla\times\left(\eta_w\nabla\times\mathbf{B}\right),
\label{eq:induction_pluto_wind}
\end{equation}
where $p$ is the total pressure (thermal and magnetic), $\mathbf{I}$ is the identity matrix, $\mathbf{a}$ is a source term (gravitational acceleration in our case), and $\eta_w$ is the resistivity in the wind domain. We use the ideal equation of state:
\begin{equation}
\rho\varepsilon = p_{th}/(\gamma -1),
\end{equation}
where $p_{th}$ is the thermal pressure, $\varepsilon$ is the internal energy per mass and $\gamma$ is the adiabatic exponent. This gives for the energy: $E = \rho\varepsilon+\mathbf{m}^2/(2\rho)+\mathbf{B}^2/(8\pi)$.

PLUTO solves normalized equations, using three quantities to set all the others: length $R_0$, density $\rho_0$ and speed $U_0$. If we note with $*$ the parameters related to the star and with $0$ the parameters related to the normalization, we have $R_*/R_0=1$, $\rho_*/\rho_0=1$ and $u_{kep}/U_0=\sqrt{GM_*/R_*}/U_0=1$, where $u_{kep}$ is the Keplerian speed at the stellar surface, $G$ the gravitational constant and $M_*$ the stellar mass. By choosing the physical values of $R_0$, $\rho_0$ and $U_0$, one can deduce all of the other values given by the code in physical units. In our set-up, we choose $R_0=R_\odot=6.96 \ 10^{10}$ cm, $\rho_0=\rho_\odot=1.68 \ 10^{-16} \ \mathrm{g/cm}^3$ (which corresponds to the density in the solar corona above 2.5 Mm, cf. \cite{vernazza_structure_1981}) and $U_0=u_{kep,\odot}=4.37 \ 10^2$ km/s. Our wind simulations are then controlled by three parameters: the adiabatic exponent $\gamma$ for the polytropic wind, the rotation of the star normalized by the escape velocity $u_{rot}/u_{esc}$ and the sound speed normalized also by the escape velocity $c_s/u_{esc}$. Note that the escape velocity is defined as $u_{esc} = \sqrt{2}u_{kep} = \sqrt{2GM_*/R_*}$. For the rotation speed, we take the solar value, which gives $u_{rot}/u_{esc} = 2.93 \ 10^{-3}$. We choose to set $c_s/u_{esc}=0.243$, which corresponds to a $1.3 \ 10^6$ K hot corona for solar parameters and $\gamma=1.05$. This choice of $\gamma$ is dictated by the need to maintain an almost constant temperature as the wind expands, which is what is observed in the solar wind. Hence choosing $\gamma \neq 5/3$ is a simplified way of taking into account heating, which is not modeled here. An evolution of this model handling the complex problem of coronal heating with Alfvén waves can be found in \cite{reville_role_2020} and \cite{hazra_modeling_2021}. We have a final control parameter $u_A/u_{esc}$ which corresponds to the Alfvén speed $u_A$ normalized by the escape velocity to control the amplitude of the initial magnetic field. The value of this parameter depends on the dynamo parameters, as explained in Appendix \ref{appendix:osc}, and thus will be presented in section \ref{sec:evol}.

\begin{figure}
    \centering
    \includegraphics[width=0.8\textwidth]{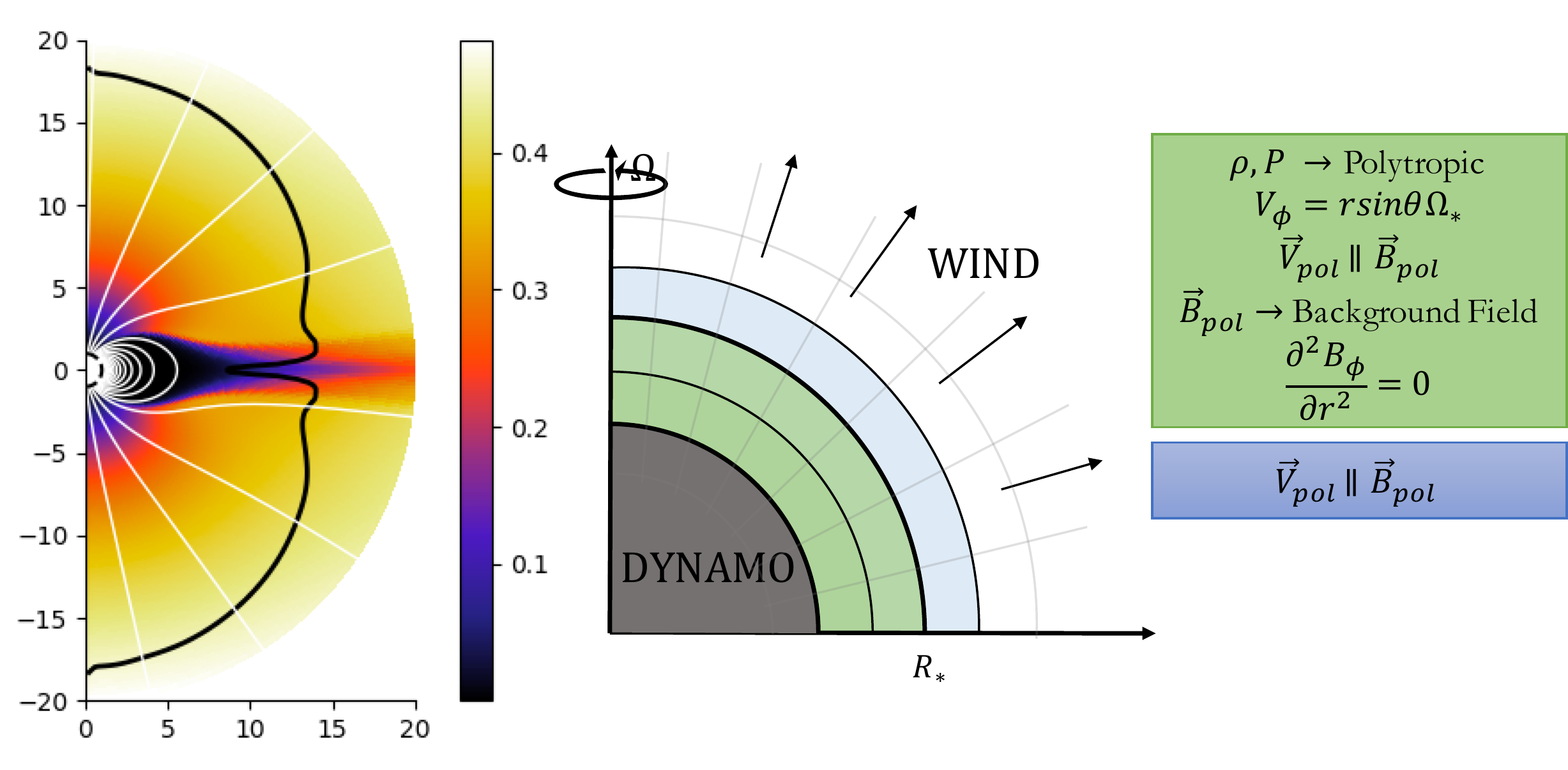}
    \caption{Presentation of the wind model. On the left is an example of the relaxed state of the corona for a dipolar magnetic field: the colorscale represents the poloidal speed in code units ($U_0=437 km/2$), the black line corresponds to the Alfvén radius and the white lines to the poloidal magnetic field lines. On the right is a scheme of the bottom boundary conditions used for the wind computational domain alone. The bottom part corresponding to the stellar interior is in grey/dark green because it is not considered yet.}
    \label{fig:wind_pic_bc}
\end{figure}

We assume axisymmetry but solve for the 3 components of flow and magnetic field. We use the spherical coordinates $(r,\theta)$. We choose a finite-volume method using an approximate Riemann Solver (here the HLL solver, cf. \cite{einfeldt_godunov-type_1988}). PLUTO uses a reconstruct-solve-average approach using a set of primitive variables $(\rho,\mathbf{u},p,\mathbf{B})$ to solve the Riemann problem corresponding to the previous set of equations. The time evolution is then implemented via a second order Runge-Kutta method. To enforce the divergence-free property of the field we use a hyperbolic divergence cleaning, which means that the induction equation is coupled to a generalized Lagrange multiplier in order to compensate the deviations from a divergence-free field \citep{dedner_hyperbolic_2002}. 

The numerical domain dedicated to the wind computation is an annular meridional cut with the co-latitude $\theta \in [0,\pi]$ and the radius $r \in [1.01,20]R_*$, as shown in the left panel of figure \ref{fig:wind_pic_bc}. We use an uniform grid in latitude with 256 points and a stretched grid in radius with 400 points, with the grid spacing geometrically increasing from $\Delta r/R_*=0.0016$ at the surface of the star to $\Delta r/R_*=0.02$ at the outer boundary. At the latitudinal boundaries ($\theta=0$ and $\theta=\pi$) we set axisymmetric boundary conditions. At the top radial boundary ($r=20 R_*$) we set an outflow boundary condition which corresponds to $\partial/\partial r=0$ for all variables, except for the radial magnetic field where we enforce $\partial(r^2B_r)/\partial r=0$. Because the wind has opened the field lines and under the assumption of axisymmetry, this ensures the divergence-free property of the magnetic field. The bottom boundary conditions that we would use with the wind computational domain only are shown in the right panel of figure \ref{fig:wind_pic_bc}. In the ghost cells (in green), the density $\rho$ and pressure $p$ are set to a polytropic profile, the rotation is uniform and the poloidal speed $U_{\rm{pol}}$ is aligned with the poloidal magnetic field $B_{\rm{pol}}$. The latter is held fixed while the toroidal magnetic field $B_\phi$ is linearly extrapolated from the numerical domain. In the first point of the computational domain (in blue) all physical quantities are free to evolve, except for the poloidal speed $U_{\rm{pol}}$ which is forced to be aligned with the poloidal magnetic field $B_{\rm{pol}}$ to minimize the generation of currents at the surface of the star and keep it as close as possible to a perfect conductor. 
We initialize the velocity field with a spherically symmetric polytropic wind solution. In the grey area we do not solve yet the dynamo, which we will now describe.

\subsection{Dynamo model}
\label{subsec:model_dyn}

\begin{figure}
    \centering
    \includegraphics[width=0.8\textwidth]{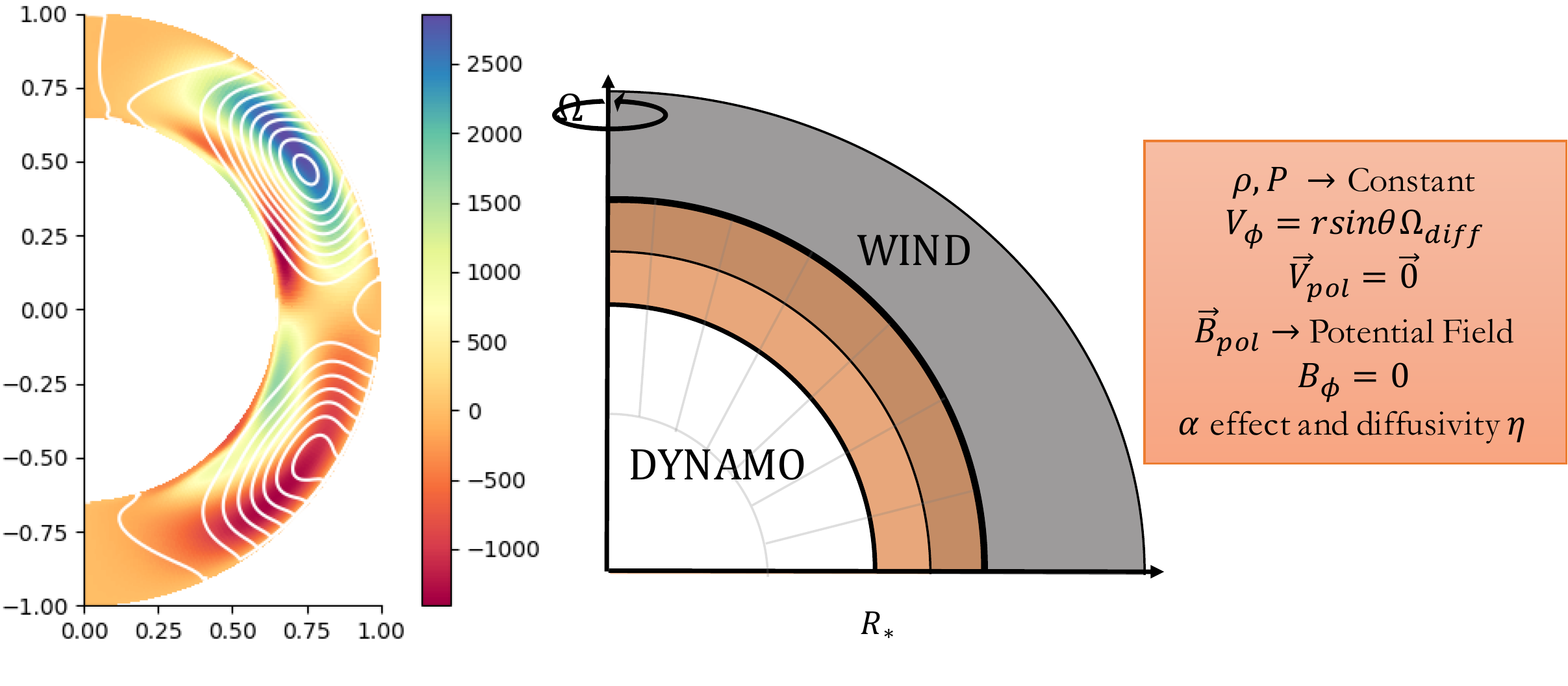}
    \caption{Presentation of the dynamo model. On the left is an example of the dynamo evolution: the colorscale represents the normalized toroidal magnetic field $B_\phi$ in code units and the white lines are the contours of the poloidal magnetic field. On the right is a scheme of the top boundary conditions used for the dynamo computational domain alone. The wind domain is in grey because it is not considered yet.}
    \label{fig:dyn_pic_bc}
\end{figure}

As we have seen in equation \ref{eq:induction_pluto_wind}, PLUTO solves the full non-linear induction equation. For the dynamo inside the star, we will consider that the magnetic field $\mathbf{B}$ is the large-scale mean field and that the velocity $\mathbf{u}_d$ is now only composed of a toroidal component $u_\phi$ (we neglect the meridional flow for now) and thus implement an alternative form of the induction equation:
\begin{equation}
\frac{\partial}{\partial t}\mathbf{B}+\nabla\cdot(\mathbf{u}_d\mathbf{B}-\mathbf{B}\mathbf{u}_d)=\nabla\times\left(\alpha\mathbf{B}\right) - \nabla\times\left(\eta\times\nabla\times\mathbf{B}\right),
\end{equation}
where $\eta$ is the effective magnetic diffusivity and $\alpha$ is a parametrized coefficient (the $\alpha$-effect, that can be obtained through the First Order Smoothing Approximation (FOSA) of the electro-motive force \citep{pouquet_strong_1976}. Hence the $\Omega$ effect is taken into account with the second term and the $\alpha$ effect with the third one. We can choose to have an $\alpha$ or $\alpha^2$ effect by disabling or enabling the third term for the equation on $B_\phi$. This form of the induction equation is only active inside the star ($r<R_*$). The other equations \ref{eq:rho_wind}, \ref{eq:m_wind} and \ref{eq:E_wind} are not solved inside the star.

This new induction equation follows the same normalization as described before. However the mean field dynamo community usually refers to the control parameters $C_\alpha$ and $C_\Omega$ obtained with a different normalization. To make it more convenient we will use in this article the traditional control parameters of the dynamo models when discussing our solution, but we have described in Appendix \ref{appendix:norm} the difference between the two and hence which conversion factor we apply in our code to respect the PLUTO normalization.

The flow in the dynamo domain will consist here only of a differential rotation such that $u_\phi = \Omega_0 r\rm{sin} \ \theta\Omega(r,\theta)$. The rotation in this zone is solar-like \citep{thompson_internal_2003} with a solid body rotation below $0.66R_*$ and differential rotation above, with the equator rotating faster than the poles (25 versus 35 days). We use the following normalized profile:
\begin{equation}
\Omega(r,\theta) = \Omega_c + \frac{1}{2}\left(1+\rm{erf}\left(\frac{r-r_c}{d}\right)\right)\left(1-\Omega_c-c_2\rm{cos}^2\theta\right),
\end{equation}
where $\Omega_c = 0.92$, $r_c=0.7R_*$, $d = 0.02R_*$ and $c_2 = 0.2$. The corresponding control number $C_\Omega=\Omega_0R_*^2/\eta_t$ is always fixed at $1.4\times10^5$ which corresponds to an equatorial rotation rate of $\Omega_0/2\pi=456$ nHz for $\eta_t=10^{11} \ \rm{cm}^2.\rm{s}^{-1}$, which is in good agreement with observation of the actual solar rotation rate. 

As a first simple approximation the $\alpha$ effect is constant in the convection zone and zero in the radiative zone below with a smooth transition between the two zones. We use the following profile studied by the benchmark of \cite{jouve_solar_2008}:
\begin{equation}
\alpha(r,\theta) = \alpha_0 \frac{3\sqrt{3}}{4}\,\rm{sin}^2\theta\,\rm{cos}\,\theta\left(1+\rm{erf}\left(\frac{r-r_c}{d}\right)\right) \frac{1}{T_{quench}}.
\end{equation}
The physical amplitude of the $\alpha$ effect is given by the $C_\alpha = \alpha_0 R_\star/\eta_t$ parameter, which value will vary. The quenching term $T_{quench}$ allows for saturation at the reference magnetic field value $B_{quench}$:
\begin{equation}
T_{quench} = 1+\left(\frac{B_\phi(r,\theta,t)}{B_{quench}}\right)^2.
\label{eq:dyn_quench}
\end{equation}

\begin{table}[t!]
\centering
\begin{tabular}{|c||c|c|} 
\hline
Case & $C_{\alpha}^{crit}$ & $\omega$ \\ \hline
$A_{bench}$ & $0.387 \pm 0.002$ & $158.1 \pm 1.472$ \\
$A_{PLUTO}$ & 0.385 & 157.1 \\ \hline
$A'_{bench}$ & $0.369 \pm 0.002$ & $157.4 \pm 0.894$ \\
$A'_{PLUTO}$ & 0.366 & 158.2 \\ \hline \hline
$B_{bench}$ & $0.408 \pm 0.003$ & $172.0 \pm 0.632$ \\
$B_{PLUTO}$ & 0.410 & 170.8 \\ \hline
$B'_{bench}$ & $0.387 \pm 0.002$ & $168.8 \pm 0.447$ \\
$B'_{PLUTO}$ & 0.392 & 168.5 \\ \hline
\end{tabular}
\caption{Comparison of the critical dynamo number $C_\alpha^{crit}$ and the cycle period $\omega$ between the benchmark described in \cite{jouve_solar_2008} and our dynamo set-up for various cases. The letters A and B refer to cases without and with a jump in diffusivity. Cases with a prime refer to radial boundary conditions, while cases without refer to potential boundary conditions.}
\label{tab:bench}
\end{table}

We have a swift transition in diffusivity between the radiative and the convection zone of two orders of magnitudes, following this profile:
\begin{equation}
\eta(r) = \eta_c + \frac{1}{2}\left(\eta_t-\eta_c\right)\left(1+\rm{erf}\left(\frac{r-r_c}{d}\right)\right),
\end{equation}
with $\eta_c = 10^9 \ \rm{cm}^2.\rm{s}^{-1}$ and $\eta_t = 10^{11} \ \rm{cm}^2.\rm{s}^{-1}$.

The magnetic field is initialized with a dipole confined in the convection zone:
\begin{equation}
A_\phi = A_0\frac{R_\star^2 \rm{sin}\,\theta}{r^2}\left(1 - \frac{1}{2}\left(1.0 + \rm{tanh}\left(\frac{r_c-r}{d}\right)\right)\right),
\end{equation}
where $A_0=u_A\sqrt{4\pi\rho_*}R_\star$. Hence $B_\phi$ is initially equal to 0.

The numerical domain dedicated to the dynamo computation is an annular meridional cut with the co-latitude $\theta \in [0,\pi]$ and the radius $r \in [0.6,1.01]R_*$, as shown in the left panel of figure \ref{fig:dyn_pic_bc}. We use a uniform grid in latitude with 256 points and a uniform grid in radius with 200 points, which yields a grid spacing of $\Delta r/R_*=0.002$. At the latitudinal boundaries ($\theta=0$ and $\theta=\pi$) we set axisymmetric boundary conditions. For the bottom boundary condition ($r=0.65R_*$) we use a perfect conductor condition:
\begin{equation}
A_\phi = 0, \frac{\partial (rB_\phi)}{\partial r} = 0.
\end{equation}
The top boundary conditions ($r=R_*$) that we use when only the dynamo model is running are shown in the right panel of figure \ref{fig:dyn_pic_bc}.
The poloidal magnetic field $B_{\rm{pol}}$ is matched to a potential field following the method described in appendix \ref{appendix:extra}, while the toroidal magnetic field $B_\phi$ is set to 0.

Since this kind of kinematic dynamo had never been implemented before in the PLUTO code, we validated our model by comparing it with the benchmark described in \cite{jouve_solar_2008}. Table \ref{tab:bench} compares the critical threshold for the dynamo number $C_\alpha^{crit}$ to have a dynamo solution growing exponentially and the period $\omega$ of the corresponding cycle obtained in our dynamo set-up to the published benchmark cases. The letters A and B refer to cases without and with a jump in diffusivity. Cases with a prime refer to radial boundary conditions while cases without refer to potential boundary conditions. Our cases show a good agreement within the error bars found in the benchmark and we are thus confident that our dynamo set-up is numerically robust.

\subsection{Wind-dynamo interface}
\label{subsec:model_interface}

\begin{figure}
    \centering
    \includegraphics[width=\textwidth]{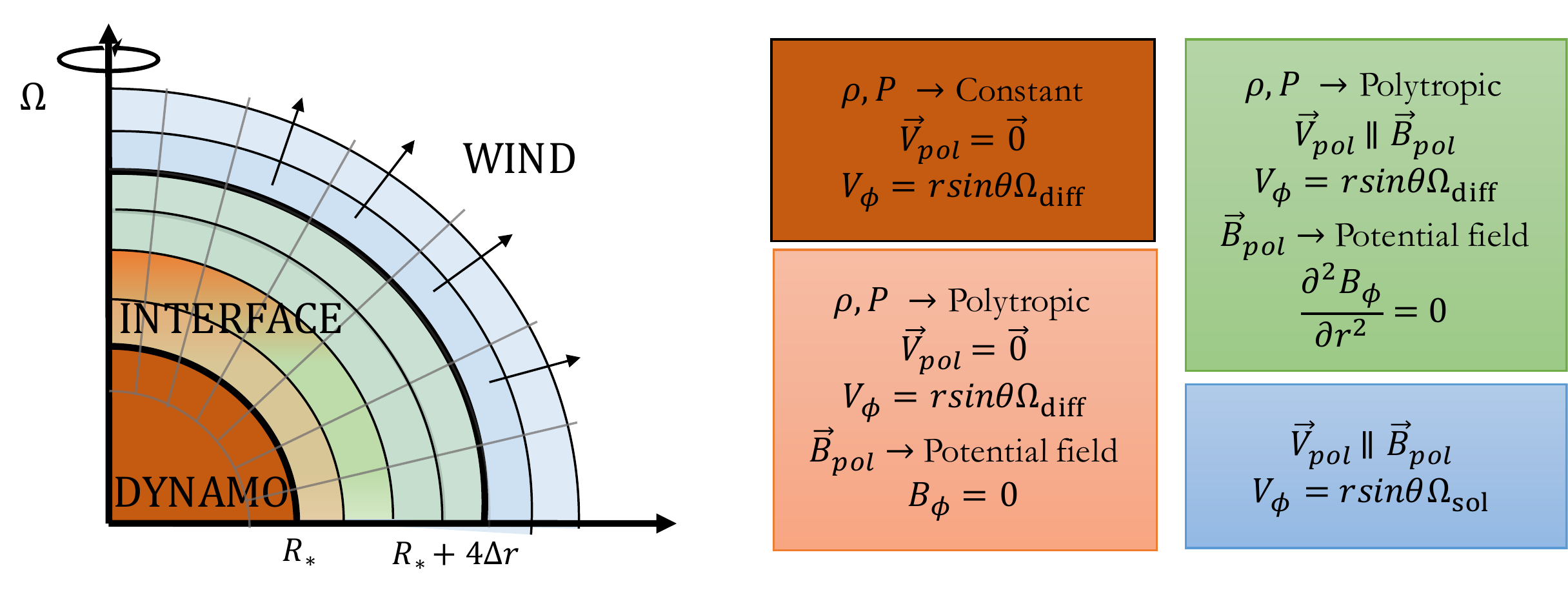}
    \caption{Description of the interface region. On the left is a scheme of the four layers between the dynamo and wind computational domains. On the right are the descriptions of the various physical quantities in each corresponding region on the scheme. There both dynamo and wind domains are jointly considered, contrary to figures \ref{fig:dyn_pic_bc} and \ref{fig:wind_pic_bc}.}
    \label{fig:bc_coupling}
\end{figure}

The interface buffer between the dynamo and the wind computational domains is shown in figure \ref{fig:bc_coupling}. This interface is crucial to control and understand the interactions between the two domains. We will describe the chosen interface in this section, combining the traditional dynamo and wind boundary conditions shown in figures \ref{fig:wind_pic_bc} and \ref{fig:dyn_pic_bc}. Now both grey zones are becoming active.

The interface is divided into four layers of one grid point each because of the 2-point stencil used for our linear reconstruction. That way, the first two layers constitute the boundary condition for the dynamo and the last two layers constitute the boundary condition for the wind. 
We also describe the first two points of the wind computational domain (blue in figure \ref{fig:bc_coupling}) where we add extra conditions for more stability. 

The first layer is very similar to the dynamo boundary conditions shown in the right panel of figure \ref{fig:dyn_pic_bc}, except that now we prescribe the density and pressure to follow a polytropic law (but continuous with the constant value inside the star). The magnetic field is extrapolated as a potential field using the value of the last point of the dynamo zone. The mathematical principle and numerical implementation of the potential field extrapolation are detailed in appendix \ref{appendix:extra}. 

In the last two layers of the interface, similar to the right panel of figure \ref{fig:wind_pic_bc}, the magnetic field is still extrapolated beyond $R_*$, the poloidal speed is aligned with the poloidal magnetic field and $\partial_r^2B_\phi$ is set to 0 to limit the growth of currents at the surface of the star.

The second layer is special because we can enforce two types of boundary condition. It can either be similar to the first layer, so closer to the dynamo boundary condition: if we do so, the wind computational region has absolutely no impact on the dynamo boundary conditions. This leads to what we call the \textit{one-way coupling} where only the dynamo can influence the wind, without feedback from the latter. On the contrary we can choose to enforce a condition similar to the last two layers, so closer to the wind boundary condition: if we do so, the wind can back-react on the dynamo boundary conditions via the $B_\phi$ condition which is derived from the wind computational domain. This is what we call the \textit{two-way coupling} where both the dynamo and the wind influence each other. 

In the first two layers of the wind computational domain, we still impose the poloidal speed to limit again the generation of currents. We also impose some diffusivity a bit further than the star surface using the following expression:
\begin{equation}
\eta_w = \frac{1}{2}\eta\left(1+\rm{tanh}\left(\frac{r_\eta-r}{d_\eta}\right)\right),
\label{eq:eta_profile}
\end{equation}
with $r_\eta = 1.015R_*$ and $d_\eta = 0.003R_*$. This allows the diffusivity to drop only after around ten grid points above the interface. As we explain in section \ref{appendix:osc}, this additional diffusivity helps for the stability of our coupled dynamo-wind model and allows the information from the dynamo domain to be smoothly transmitted to the wind domain.

The final simulation box obtained is thus a combination of a dynamo computational domain, an interface buffer region and a wind computational domain, as can be seen in figure \ref{fig:dp_snap_1w}. Thus the poloidal magnetic field generated by the dynamo effect inside the star can impact directly the co-rotating idealized corona, with propagating timescales characterized in Appendix \ref{appendix:osc}.

\section{Consistent evolution of dynamo and wind quantities along an activity cycle}
\label{sec:evol}

To validate the coupling from a theoretical point of view we focus on a model that develops a fast magnetic cycle, thus probably more related to a young Sun with the same rotation rate but enhanced activity. This allows us to reduce the discrepancy between the dynamo and the wind timescales and thus compute the evolution along several activity cycles using available computation resources. The main characteristics of this model, which we will now refer to as DW (for Dynamo-Wind), are given in table \ref{tab:dw}. We use the magnetic diffusivity to set the cycle period by adjusting the diffusive time $t_\eta = R_*^2/\eta_t$. Here it corresponds to a 5-day period for the magnetic cycle. Then we set $C_\Omega$ to have the solar rotation rate and finally we set $C_\alpha$ to have the proper dynamo number $D=C_\Omega \times C_\alpha$ to trigger the dynamo. This yields a case where $C_\Omega$ is set to a smaller value than $C_\alpha$ (which is less frequent than the reverse), which corresponds to a strong generation of poloidal field due to turbulence. Because of this parameter range we initialize the magnetic field with a small value to prevent growth up to unrealistic values, with a value of $u_A/u_{esc}$ of 0.05.

\begin{table}[!h]
\centering
\begin{tabular}{|c||c|c|c|c|c|c|c|c|}
\hline
Name & $\eta_t \ (\rm{cm}^2.\rm{s}^{-1})$ & $C_\alpha$ & $C_\Omega$ & $B_0$ (G) & $u_a/u_{esc}$ & $\gamma$ & $c_s/u_{esc}$ & $\rho_*$  \\ \hline\hline
DW & $3.7 \times 10^{14}$ & $1.44\times10^4$ & $3.4\times10^1$ & $3.0\times10^{-3}$ & $0.05$ & $1.05$ & $0.243$ & $0.55$ \\ \hline
\end{tabular}
\caption{Table of the main parameters for the test case DW.}
\label{tab:dw}
\end{table}

Another important parameter to set is the quenching value of the toroidal field $B_{quench}$. It is indeed crucial to 
ensure that the large-scale magnetic field does not reach unrealistic values of $u_A/u_{esc}$ (see \cite{reville_solar_2015}). We have thus derived an empirical criterion to stabilize the poloidal maximum surface energy close to the input value of the simulation. To do so, we have run 4 simulations with the same dynamo number $D$ but with different ratios of $C_\alpha$ and $C_\Omega$, and have found the following relationship:
\begin{equation}
B_{quench} = 0.17 B_{pol}^{surf}\sqrt{\frac{C_\alpha}{C_\Omega}}.
\end{equation}
We can then use our input parameters to compute an estimate of $B_{quench}$, given a desired surface poloidal field.

We will first present a one-way coupling version of the DW model, which we will call DW1. This allows us to show the impact of the dynamo-generated magnetic field on the corona. For the first time we obtain a self-consistent solution using one simulation domain. This model is indeed already a great improvement in regards of the quasi-static methods \citep{perri_simulations_2018} because it allows for a dynamical influence of the dynamo magnetic field on the wind without any risk of breaking causality (given that we respect the criteria derived in appendix \ref{appendix:osc}).

\begin{figure}
    \centering
    \includegraphics[width=\textwidth]{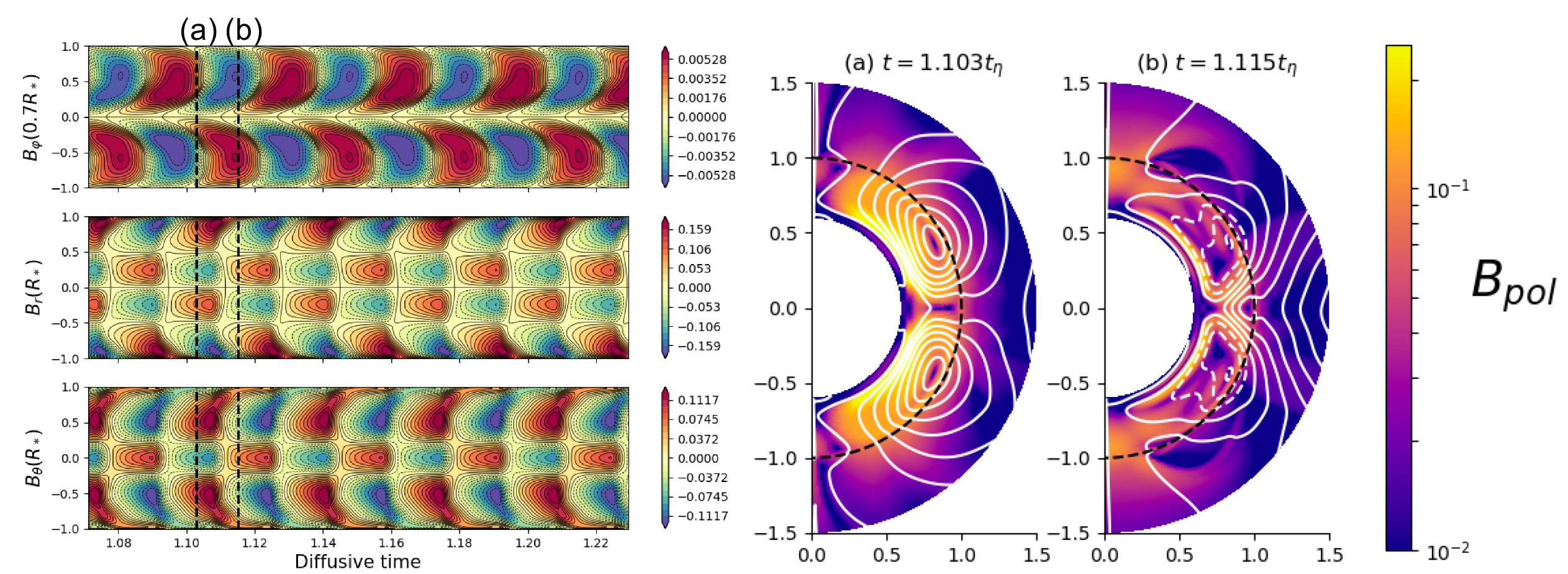}
    \caption{Left panel: Butterfly diagram of the dynamo solution for case DW1. We show the toroidal magnetic field $B_\phi$ at the base of the convective zone ($r=0.7R_*$) in the top panel, the radial magnetic field $B_r$ at the star surface in the middle panel, and the latitudinal magnetic field $B_\theta$ at the surface in the bottom panel, all in Gauss units. \\ Right panel: Meridional cuts of the dynamo-wind solution at different times for case DW1. We show the norm of the poloidal magnetic field over the first 1.5 solar radii, in code units. Magnetic field lines are in white (full line when clockwise, dashed when counter clockwise). The surface of the star is indicated as a black dashed line. The corresponding times are shown as black dashed lines on the butterfly diagram in the left panel.}
    \label{fig:dp_snap_1w}
\end{figure}

We start by presenting the dynamo solution. The butterfly diagram obtained through this one-way coupling is presented in the left panel of figure \ref{fig:dp_snap_1w}. We have chosen to display the evolution between approximately $1.07$ and $1.23$ $t_\eta$ to avoid initial transients and focus on typical structures. The top panel shows the toroidal magnetic field $B_\phi$ at the base of the tachocline ($0.7R_*$), the middle panel shows the radial magnetic field $B_r$ at the surface of the star and the bottom panel shows the latitudinal magnetic field $B_\theta$ also at the surface of the star. They are all in Gauss units.
At the base of the tachocline (top panel), we can clearly see the reversal of polarity of the toroidal magnetic field, along the cyclic evolution of the surface poloidal field into as well (bottom panels). The period of the dynamo cycle is approximately 0.04 $t_\eta$. The cycle displays both equatorial and polar branches at the surface due to our choice of $\alpha$ effect. It corresponds to a so-called Parker-Yoshimura dynamo wave \citep{yoshimura_model_1975}.
Over the temporal range considered the cyclic magnetic field is always anti-symmetric with respect to the equator. 
The right panel of figure \ref{fig:dp_snap_1w} shows meridional cuts taken at moments of interest during the evolution of the solution. We can see the norm of the poloidal magnetic field over the first 1.5 solar radii, in code units ($B_0 = 2$ G). The surface of the star is marked by a black dashed line while the magnetic field lines are displayed in white (full line when clockwise, dashed when counter clockwise). Snapshot (a) allows us to see the regeneration of the poloidal magnetic field due to the dynamo effect inside the star, more precisely at the tachocline ($0.7R_*$) due to our distribute $\alpha$-effect. The magnetic structure appears mostly dipolar at large-scale, but more octupolar closer to the surface (both of anti-symmetric parity). We can already see the field lines crossing the interface continuously into the corona. Snapshot (b) shows then the advection of the magnetic structures and their transmission to the corona thanks to our potential field boundary condition. Once in the corona the structures interact with the stellar wind which tends to open the field lines and quickly decrease in amplitude. We begin to see the regeneration of the next cycle structures inside the tachocline, with the opposite polarity. The corresponding times are shown as black dashed lines on the butterfly diagram in the left panel.

\begin{figure}
    \centering
    \includegraphics[width=\textwidth]{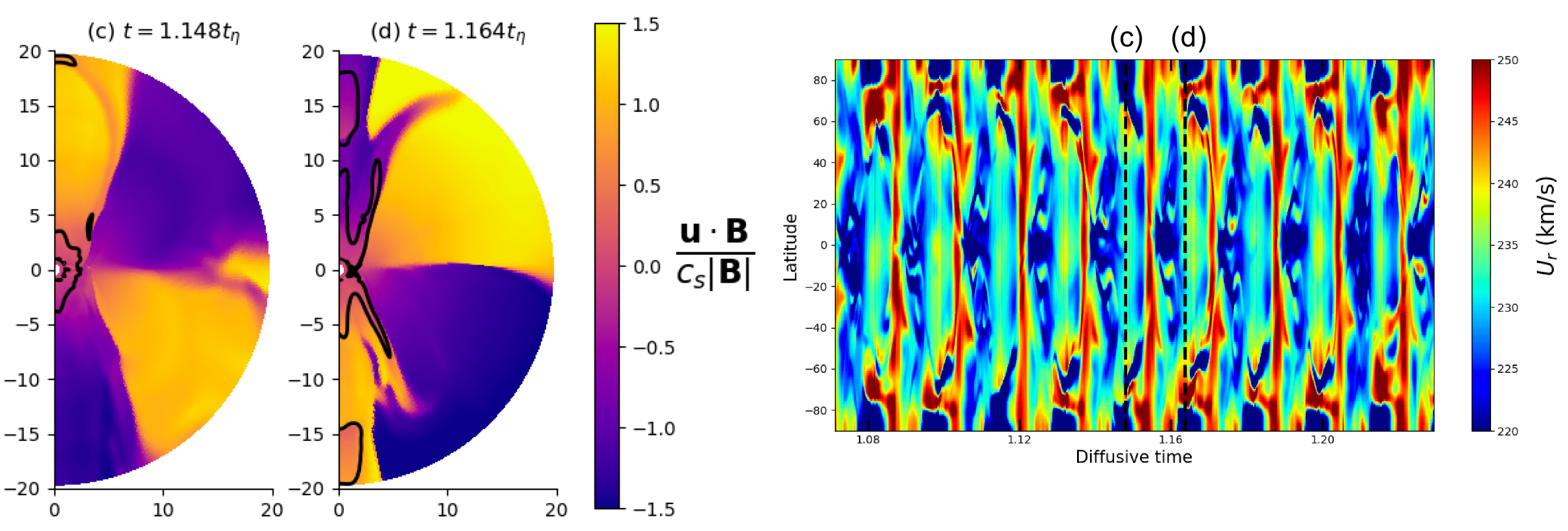}
    \caption{Left panel: Meridional cuts of the dynamo-wind solution at two different times for case DW1. We show the quantity $\mathbf{u}\cdot\mathbf{B}/(c_S|\mathbf{B}|)$ which corresponds to the Mach number projected on the normalized magnetic field. The black line is the Alfvén surface which is the distance at which the wind speed becomes equal to the Alfvén speed. The surface of the star is indicated as a black dashed line. We clearly notice various polarity sectors. \\ Right panel: Time-latitude diagram of the wind solution for case DW1. We show the wind speed $u_r$ at the far end of our computational box ($r=20R_*$) in km/s. The corresponding times for the meridional cuts in the left panel are shown as black dashed lines on the wind diagram.}
    \label{fig:dw_snap_1w}
\end{figure}

Now we will present the response of the corona to this cyclic dynamo for the one-way coupling.
The left panel of figure \ref{fig:dw_snap_1w} shows meridional cuts taken at moments of interest during the evolution of the DW1 solution. We can see the Mach number projected on the normalized magnetic field up to 20 $R_*$. This allows us to see at the same time the various streams of the stellar wind and the corresponding polarity of the magnetic field associated. The black line shows the Alfvén surface which corresponds to the distance at which the wind speed becomes equal to the Alfvén speed. The surface of the star is marked by a black dashed line. We can see that the corona is highly dynamical with transients generated at the surface of the star and carried by the wind to the external boundary of the computational domain. Due to the cyclic nature of the dynamo, the hemispheres reverse in polarity. In snapshot (c) the northern hemisphere is divided between two zones, one positive near the pole and one negative near the equator. In snapshot (d) however, we have a disposition similar but with reverse polarity: the coronal hole near the pole is now negative and the equator is positive. This behavior is the same with opposite polarity in the southern hemisphere. Finally we can notice that the Alfvén surface is displaying various shapes along the cycle, which is a marker of the topological changes of the dynamo field passing into the corona. Occasionally this surface can come very close to the star surface. This is due to both the low value of the magnetic field that we have to maintain with the quenching to ensure the correct coupling of the stellar interior and exterior, and also the dynamical evolution of the dynamo which can generate localized region of very low-amplitude surface field.
We now turn to studying the long-term evolution of the corona by considering the time-latitude diagram of the wind speed in the right panel of figure \ref{fig:dw_snap_1w}. The radial velocity $u_r$ is taken at the outer boundary of the domain ($20R_*$) and is shown in km/s. The corresponding times for the meridional cuts in the left panel are shown as black dashed lines on the wind diagram. We observe latitudinal variations as a response to the coupling with the dynamo-generated magnetic field. They also present themselves as cyclic with a period which is half of the dynamo cycle (around $0.02t_\eta$ vs $0.04t_\eta$), because it is dictated by the topological changes rather than the polarity changes. We observe an alternation of slow (in blue) and fast (in red) streams at the poles and at the equator. Because of the rapidly changing magnetic field, we also observe regimes where fast streams are at all latitudes while the corona is adapting to the stellar interior. This is reminiscent of the solar wind observed during the minimum and maximum of the solar cycle \citep{mccomas_three-dimensional_2003}: we don't have the same structures because we do not model the Sun but we still achieve a coupling between the dynamo field and the wind structures. The speeds reached are between 200 and 270 km/s at 20 $R_*$, which corresponds to between 400 and 520 km/s if we extrapolate them at 1 AU. Thus we recover mostly the slow component of the wind, which is typical of polytropic winds and is expected given our choice of $c_s/u_{esc}$. 

\begin{figure}
    \centering
    \includegraphics[width=\textwidth]{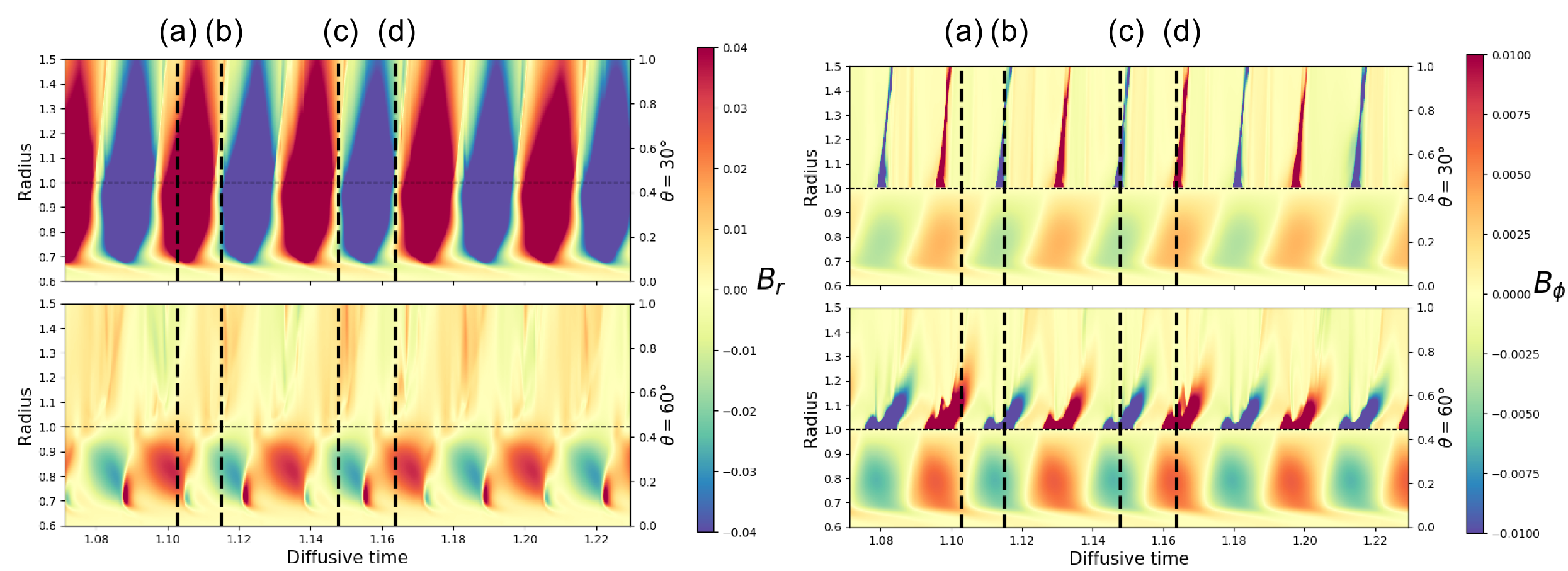}
    \caption{Time-radius diagrams of $B_r$ and $B_\phi$ (in code units, $B_0=2$ G) for case DW1 at different latitudes: the top panel is at co-latitude $\theta=30\degree$ (closer to the northern pole), the bottom panel at co-latitude $\theta=60\degree$ (mid-latitude in the northern hemisphere). The stellar surface is marked by a dashed black line at $1 R_*$. We only show the radial evolution from $0.6 R_*$ to $1.5 R_*$ to focus on the most intense structures. The times corresponding to the meridional cuts shown in figures \ref{fig:dp_snap_1w} and \ref{fig:dw_snap_1w} are shown as black dashed lines on the time-radius diagrams with the associated letter.}
    \label{fig:rt_diag_1w}
\end{figure}

We now want to understand how the one-way coupling operates. Figure \ref{fig:rt_diag_1w} shows time-radius diagrams of $B_r$ and $B_\phi$ at two latitudes: the top panel is at co-latitude $\theta=30\degree$ (closer to the northern pole), the bottom panel at co-latitude $\theta=60\degree$ (mid-latitude in the northern hemisphere). This allows us to focus on the transmission of the magnetic field from the stellar interior to the corona, with the stellar surface marked by a black dashed line at $1.0 R_*$. The times corresponding to the meridional cuts shown in figures \ref{fig:dp_snap_1w} and \ref{fig:dw_snap_1w} are shown as black dashed lines on the time-radius diagrams with the associated letter. We can clearly see the magnetic structures being generated at the base of the convection zone at $0.7 R_*$, then being advected to the stellar surface and finally being transmitted to the corona with the potential field boundary condition for $B_r$ on the left panel. The structures then are influenced by the solar wind to the end of the computational domain but also quickly decrease in intensity. We can also see the reversals in polarity due to the cyclic dynamo which translates into reversals of polarity in the corona as well, but with a delay due to the transmission time. The various latitudes show that the continuity of the stellar corona is different depending on the strength and geometry of the field. For example, $\theta=60\degree$ corresponds to less intense dynamo structures which decrease rapidly at the surface of the star, so the transmission leads to less intense structures in the wind. For $B_\phi$ we can see clearly the discontinuity created by our interface: the dynamo converges to a zero intensity toroidal field at the surface while the wind needs to have a stellar surface with small electric currents. This leads however to the periodic generation of toroidal magnetic structures, which happen at the same time that the reversal in polarity for $B_r$. It is thus very likely a result of the adjustment of the corona because of our condition on $B_\phi$ and the necessity to keep a divergence-free field in the domain. In this case this leads to the fact that the polarity of the structures of $B_\phi$ in the corona is synchronized with the one from the interior.
Here in the one-way coupling there is no feedback from the wind, so the dynamo itself is not different from a model without atmosphere.

To complement our analysis of the DW1 case, we can further compute some key quantities of our stellar dynamo model. 

We start with the magnetic characteristic length scale $l_b$ defined as:
\begin{equation}
l_b = \frac{\sum_\ell \ell \beta_{\ell,0}^2}{\sum_\ell \beta_{\ell,0}^2},
\end{equation}
where $B_r(r=R_*,\theta) = \sum_{\ell=1}^{\ell_{max}} \beta_{\ell,0}Y_\ell^0(\theta)$, so that the coefficients $\beta_{\ell,0}$ correspond to the decomposition of the surface radial magnetic field on the spherical harmonics. We only use the $m=0$ modes because in our 2.5D simulations we consider only axisymmetric fields. This length scale allows us to characterize the dominant scale in the magnetic energy power spectrum. 

We can also check the energy associated to a certain mode of the surface radial field, defined as follows:
\begin{equation}
f_{\ell} = \frac{\beta_{\ell,0}^2}{\sum_{\ell=1}^{\ell_{max}}\beta_{\ell,0}^2}.
\label{eq:fdip}
\end{equation}

\begin{figure}[!t]
    \centering
    \includegraphics[width=0.7\textwidth]{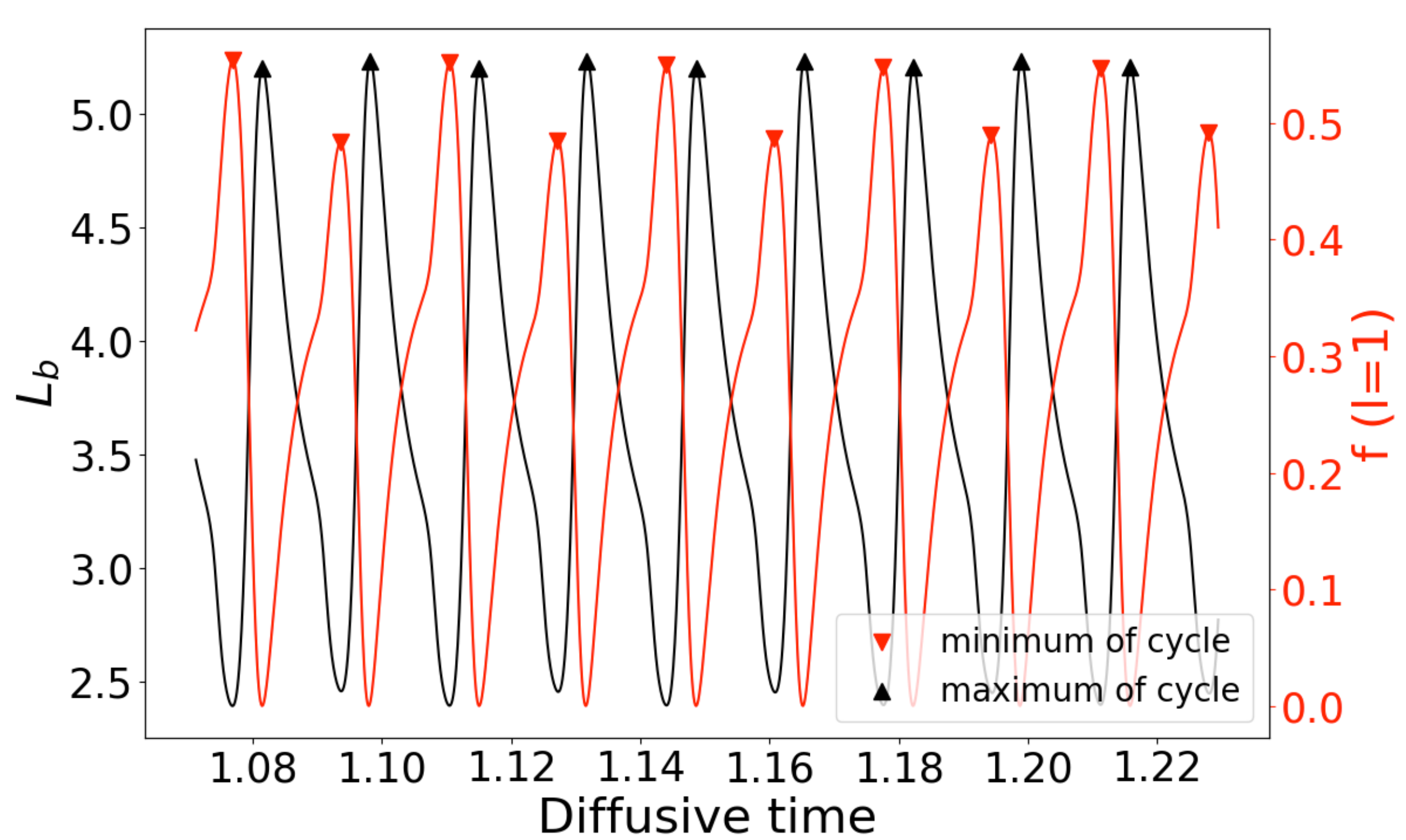}
    \caption{Time evolution of the magnetic characteristic length scale $l_b$ (in black) and the dipole coefficient $f_1$ (in red) for case DW1. Minima of activity are marked using red downward triangles, derived from $f_1$, and maxima with black upward triangles, derived from $l_b$.}
    \label{fig:lb_fdip_1w}
\end{figure}

We will use these two quantities to characterize the minima and maxima of activity in our dynamo solution.

The time evolutions of the magnetic characteristic length $l_b$ (black)
and the fraction of energy in the dipolar component $f_1$ (in red) are shown in figure \ref{fig:lb_fdip_1w}. The characteristic length $l_b$ is evolving between 2.5 and 5, showing that the surface magnetic energy of the star is best described by the first 5 modes. The maxima of $l_b$ will correspond to the maxima of the activity cycle because these are the moments where the surface magnetic energy is at its peak value. The dipole coefficient $f_1$ evolves between $10^{-6}$ and 50\%, which means that the dynamo cycle oscillates between states where the energy is mostly contained in the dipolar mode and states where the configuration is mostly multipolar just like the Sun \citep{derosa_solar_2012}. The maxima of $f_1$ will thus correspond to the minima of the activity cycle because these are the moments when the magnetic geometry is the most dipolar. We can observe an anti-correlation between the two quantities just like what is observed for the Sun \citep{derosa_solar_2012}. However the minima of activity derived from $f_1$ are in phase-quadrature with the maxima of activity derived from $l_b$, which is a characteristic of this dynamo model.

We can compare the evolution of integrated quantities to what has been found in quasi-static studies \citep{pinto_coupling_2011,reville_global_2017,perri_simulations_2018}. We will focus on the average Alfvén radius $\langle r_A\rangle$ and the mass loss $\dot{M}$. 
The Alfvén radius is defined as the distance at which the wind speed equals the Alfvén speed $v_A=B/\sqrt{4\pi\rho}$. The average Alfvén radius, as defined in \cite{pinto_coupling_2011}, corresponds to the Alfvén radius averaged by the mass flux through the surface of a sphere:
\begin{equation}
\langle r_A \rangle = \frac{\int_0^\pi r^2sin\theta\rho u_r r_A(\theta)d\theta}{\int_0^\pi r^2sin\theta||\rho \mathbf{u}||d\theta}.
\label{eq:ra}
\end{equation}

The mass loss is defined as follows:
\begin{equation}
\dot{M} = 2\pi R_s^2\int_0^\pi\rho u_rsin\theta d\theta,
\label{eq:ml}
\end{equation}
where $R_s$ is the radius at which the mass loss is computed by integration through a spherical surface. In theory, when a stationnary state is reached the mass loss should be independent of $R_s$, as long as it is big enough to include all the closed magnetic structures. In our simulations, we perform an average of the mass losses computed with $R_s\in[9;20]R_*$ to have a more precise value.

\begin{figure}
    \centering
    \includegraphics[width=\textwidth]{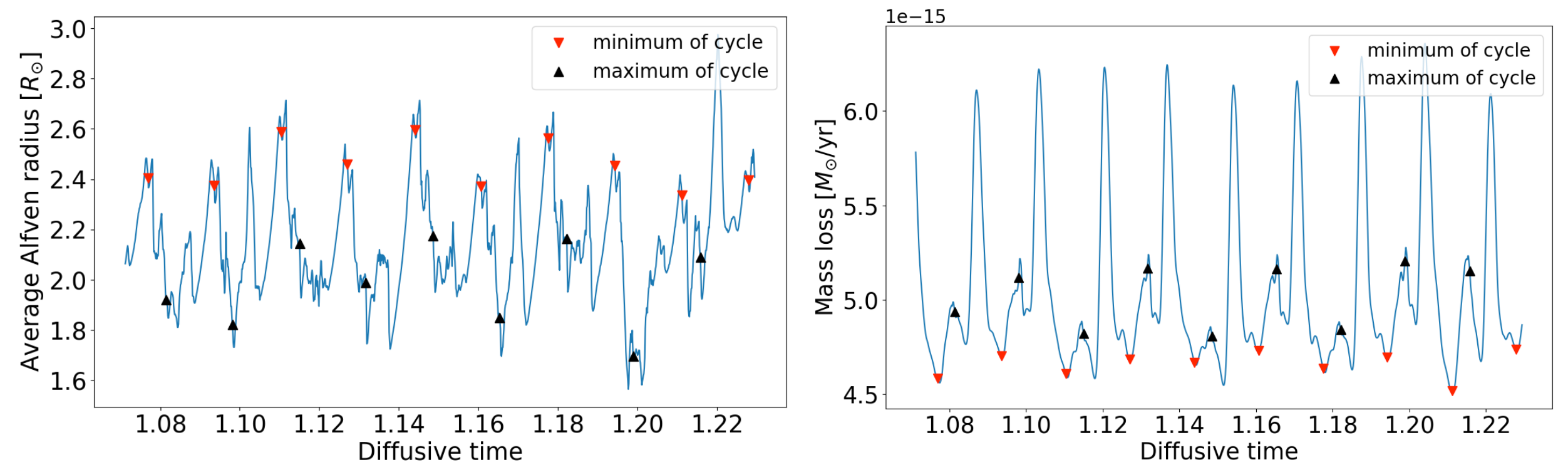}
    \caption{Time evolution of the integrated quantities of the wind for the case with one-way coupling. From left to right: the average Alfvén radius $\langle r_A\rangle$ in $R_*$ and the mass loss $\dot{M}$ in $M_\odot/\rm{yr}$. Minima of activity are marked using red downward triangles, derived from $f_1$, and maxima with black upward triangles, derived from $l_b$.} 
    \label{fig:ra_ml_1w}
\end{figure}

The temporal evolution of these quantities is shown in figure \ref{fig:ra_ml_1w}. We have added the minima and maxima of activity as defined previously by $l_b$ and $f_1$ with the same symbols as before: minima are marked using red downward triangles and maxima with black upward triangles. The average Alfvén radius varies between 1.5 and 3.0 $R_*$. This is smaller than what was found in all the previous studies cited above, but this is mainly because we did not try to use solar parameters and because of the criterion \ref{eq:osc_crit1} (see appendix \ref{appendix:osc} for more details) which limits us to weaker magnetic fields. The mass loss varies between 4.5 and 6.4 $10^{-15} \ M_\odot/\rm{yr}$, which corresponds to a 42\% variation. For comparison, the Sun has a variation of about 35\%
\citep{mccomas_weaker_2008, reville_global_2017, finley_solar_2019}. 

In this study, we are more interested in the relative variations of these quantities than their values. We see modulations in time with the evolution of the activity cycles, obtained for the first time in a completely self-consistent way. Based on similar studies, at minimum of activity we expect a bigger Alfvén radius but a smaller mass loss. At maximum of activity, it is the opposite trend. The large-scale modulations are mostly in agreement with the cycles. At minimum of activity, when the field is mostly dipolar, we indeed see a higher Alfvén radius and a lower mass loss. The maximum of activity is slightly less correlated with the integrated quantities, but most of the time it still does correspond to a smaller Alfvén radius and a bigger mass loss. We also see many small-scale variations corresponding to fluctuations and transients, which explains why the correlation is not always perfect. This may be due to the fact that we have a rapidly changing magnetic field at the surface of the star which means the wind is constantly adjusting to the magnetic configuration. 

\section{Feedback-loop between dynamo and wind}
\label{sec:feedback_loop}

We will now present what happens when we enable the feedback of the wind, which means that the wind is now able to back-react on the dynamo via the second layer of the interface and its condition on $B_\phi$ (see figure \ref{fig:bc_coupling}). We will run the same model as presented in table \ref{tab:dw} but with full two-way coupling, so now we call this two-way coupling model DW2. We will begin by showing the evolution of the global quantities as was done for DW1 in the last section. Then we will analyse the differences with the one-way coupling and quantify them in terms of dynamo modes and helicity.

\subsection{Impact of two-way coupling on global quantities}
\label{subsec:feedback_global}

\begin{figure}
    \centering
    \includegraphics[width=\textwidth]{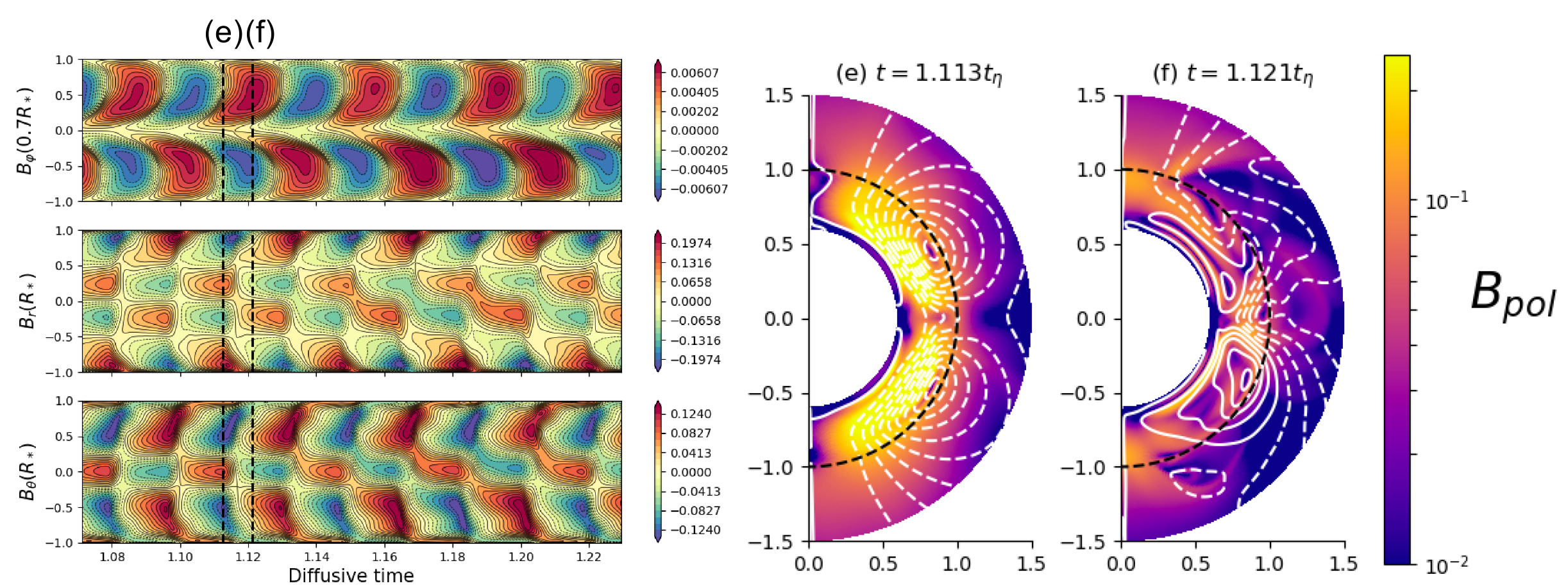}
    \caption{Left panel: Butterfly diagram of the dynamo solution for case DW2. We show the toroidal magnetic field $B_\phi$ at the base of the convective zone ($r=0.7R_*$) in the top panel, the radial magnetic field $B_r$ at the star surface in the middle panel, and the latitudinal magnetic field $B_\theta$ in the bottom panel, all in Gauss units. \\ Right panel: Meridional cuts of the dynamo-wind solution at different times for case DW2. We show the norm of the poloidal magnetic field over the first 1.5 solar radii, in code units. Magnetic field lines are in white (full line when positive, dashed when negative). The surface of the star is indicated \textbf{as} a black dashed line. The corresponding times are shown as black dashed lines on the butterfly diagram in the left panel.}
    \label{fig:dp_snap_2w}
\end{figure}

The corresponding butterfly diagram for the two-way coupling DW2 case is displayed in the left panel of figure \ref{fig:dp_snap_2w} with the same disposition as in figure \ref{fig:dp_snap_1w}. This new butterfly diagram is slightly different from the DW1 case. The toroidal magnetic field at the tachocline tends now to be more symmetrical with respect to the equator, starting at $1.12t_\eta$, indicating that the quadrupolar modes (symmetric with respect to the equator) are dominating the dynamics of the dynamo. At the surface the radial magnetic field is showing reversals of polarity but in an irregular way. It now presents north-south asymmetry most of the time and is alternating between symmetrical and anti-symmetrical configurations, which suggests an effective coupling of the dynamo families \citep{derosa_solar_2012}. The same effect is present as well for $B_\theta$. This asymmetry is highlighted by the snapshots in the right panel of figure \ref{fig:dp_snap_2w}. With the same presentation as in figure \ref{fig:dp_snap_1w}, we can see in snapshot (e) a more symmetrical configuration when the poloidal field is generated at the tachocline. But when the field rises in snapshot (f), we see that the magnetic field lines crossing the surface and influencing the corona have an asymmetric structure. The field exhibits more deformed closed loops in the northern hemisphere because of the influence of the wind than in the southern hemisphere. The magnetic structures inside the star are also not symmetrical, with the northern hemisphere rising faster than the southern one. The corresponding times are shown as black dashed lines on the butterfly diagram in the left panel.

\begin{figure}
    \centering
    \includegraphics[width=0.7\textwidth]{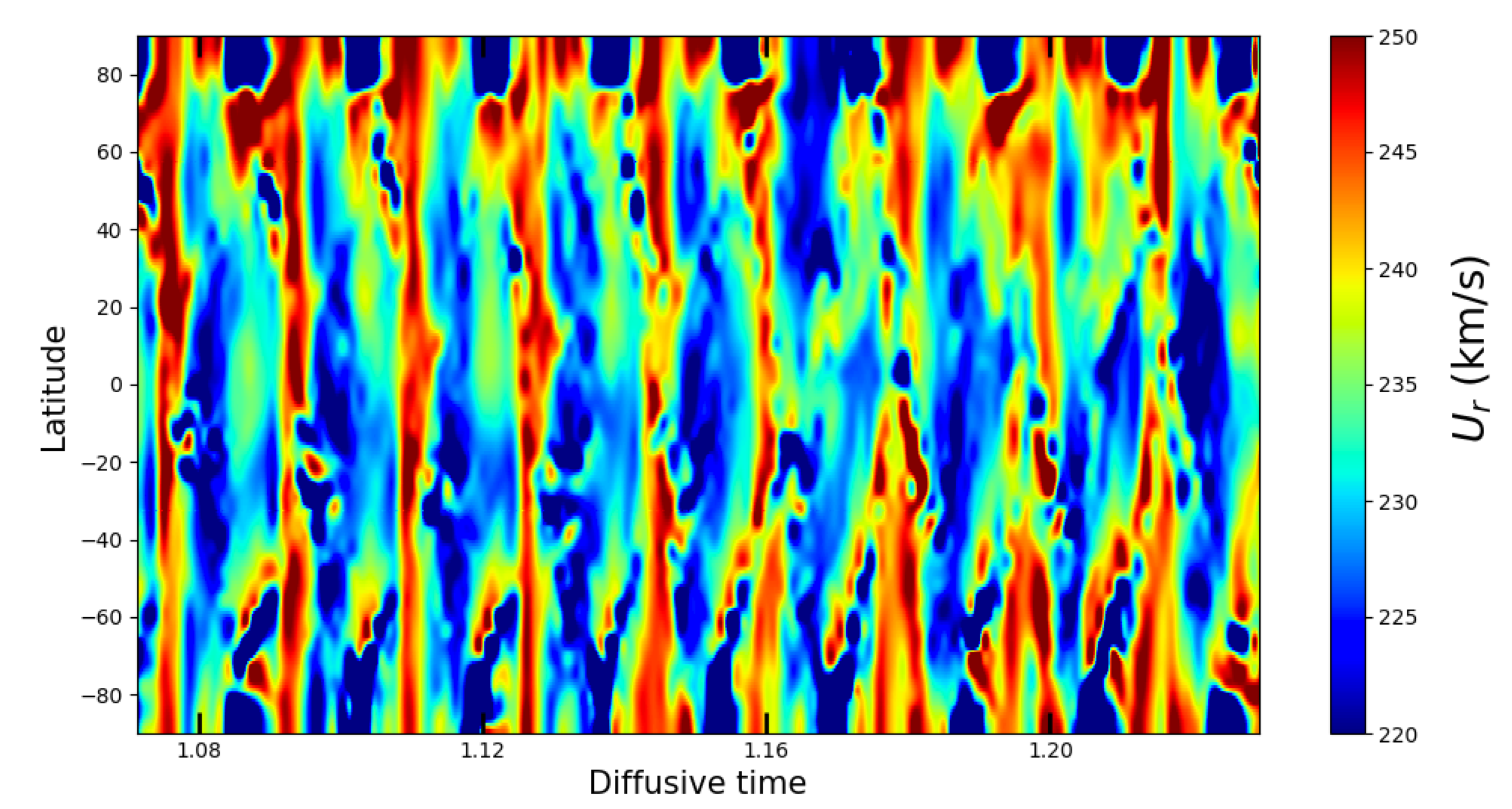}
    \caption{Time-latitude diagram of the wind solution for case DW2. We show the wind speed $u_r$ at the far end of our computational box ($r=20R_*$) in km/s. We clearly see the alternance of slower and faster wind streams, and a huge degree of asymmetry compared to figure \ref{fig:dw_snap_1w}.}
    \label{fig:dw_2w}
\end{figure}

In figure \ref{fig:dw_2w}, we present the corresponding time-latitude diagram of the radial wind velocity $u_r$ for the two-way coupling model. We do not display the corresponding meridional snapshots of the corona like in figure \ref{fig:dw_snap_1w} because the evolution of the corona is very similar, apart from the north-south asymmetry we have already reported. The evolution of the corona is still highly dynamical with transients. When the butterfly diagram is more symmetrical at the surface (for example between 1.15 and 1.25 $t_\eta$), we can see that the coronal structure becomes more irregular: there is a big patch of slow wind at the northern pole at 1.17 $t_\eta$ which is not reproduced elsewhere; fast streams seem also to be more dominant in the southern hemisphere over this period of time considered. This shows already that the feedback is introducing more variability in both the dynamo and the associated wind. 

\begin{figure}
    \centering
    \includegraphics[width=\textwidth]{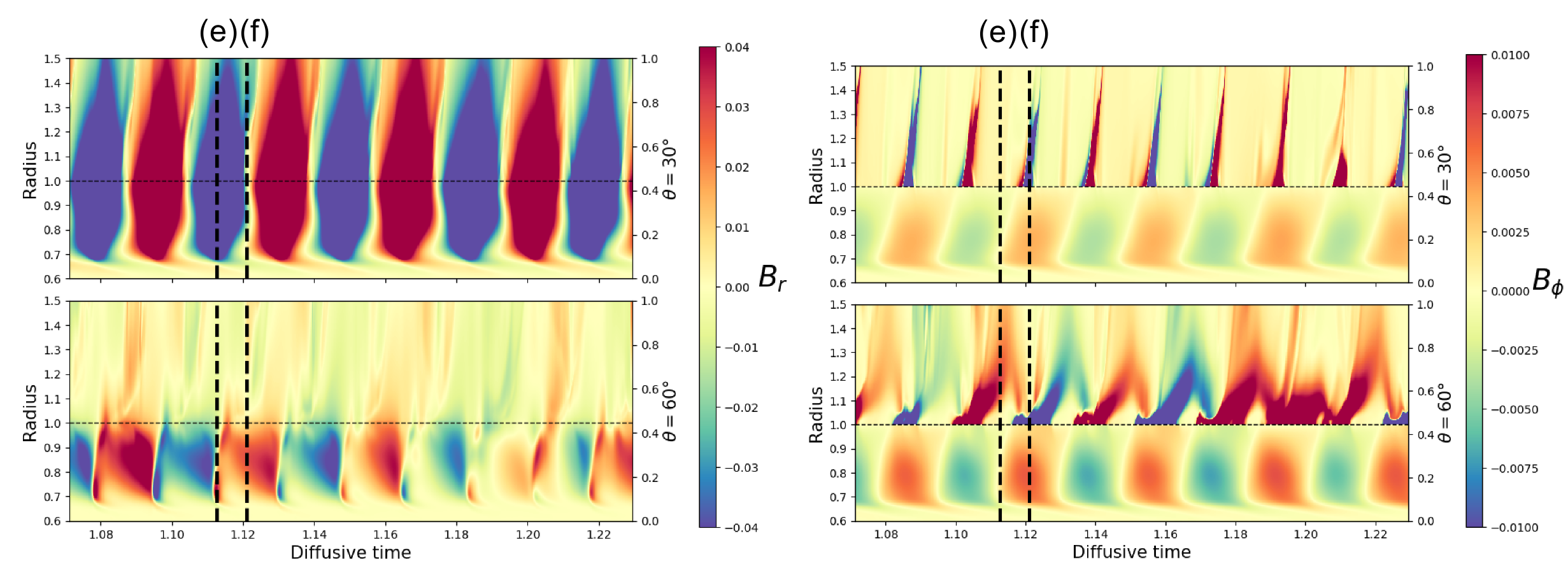}
    \caption{Time-radius diagrams of $B_r$ and $B_\phi$ (in code units, $B_0=2$ G) for case DW2 at different latitudes: the top panel is at co-latitude $\theta=30\degree$ (closer to the northern pole), the bottom panel at co-latitude $\theta=60\degree$ (mid-latitude in the northern hemisphere). The stellar surface is marked by a dashed black line at $1 R_*$. We only show the radial evolution from $0.6 R_*$ and $.1.5 R_*$ to focus on the most intense structures. The times corresponding to the meridional cuts shown in figure \ref{fig:dp_snap_2w} are shown as black dashed lines on the time-radius diagrams with the associated letter.}
    \label{fig:rt_2w}
\end{figure}

Figure \ref{fig:rt_2w} shows time-radius diagrams of $B_r$ and $B_\phi$ for case DW2 with two-way coupling. The times corresponding to the meridional cuts shown in figure \ref{fig:dp_snap_2w} are shown as black dashed lines on the time-radius diagrams with the associated letter. Once again, we clearly see the alternance of polarity and the efficient transmission of $B_r$ at $\theta=30\degree$. We can also see that the feedback of the wind has modified the dynamo boundary conditions, so that at $\theta=60\degree$ the magnetic structures do not decrease as much in intensity. For $B_\phi$, we are now left with only one point where $B_\phi=0$, the other points depend on the dynamics of the wind.
This results in more continuous and extended radial magnetic field structures at the star surface, especially at $\theta=60\degree$. These structures are once again correlated with the changes of polarity of $B_r$ in the corona. In case DW2, this leads to an opposition of polarity between the structures in the star and those in the corona. This is due to the specific period we have selected which focuses on a time with visible north-south asymmetry. We have verified that at different times when the quadrupole modes are weaker, the polarity of $B_\phi$ over the interface matches as in figure \ref{fig:rt_diag_1w}.

\begin{figure}
    \centering
    \includegraphics[width=0.7\textwidth]{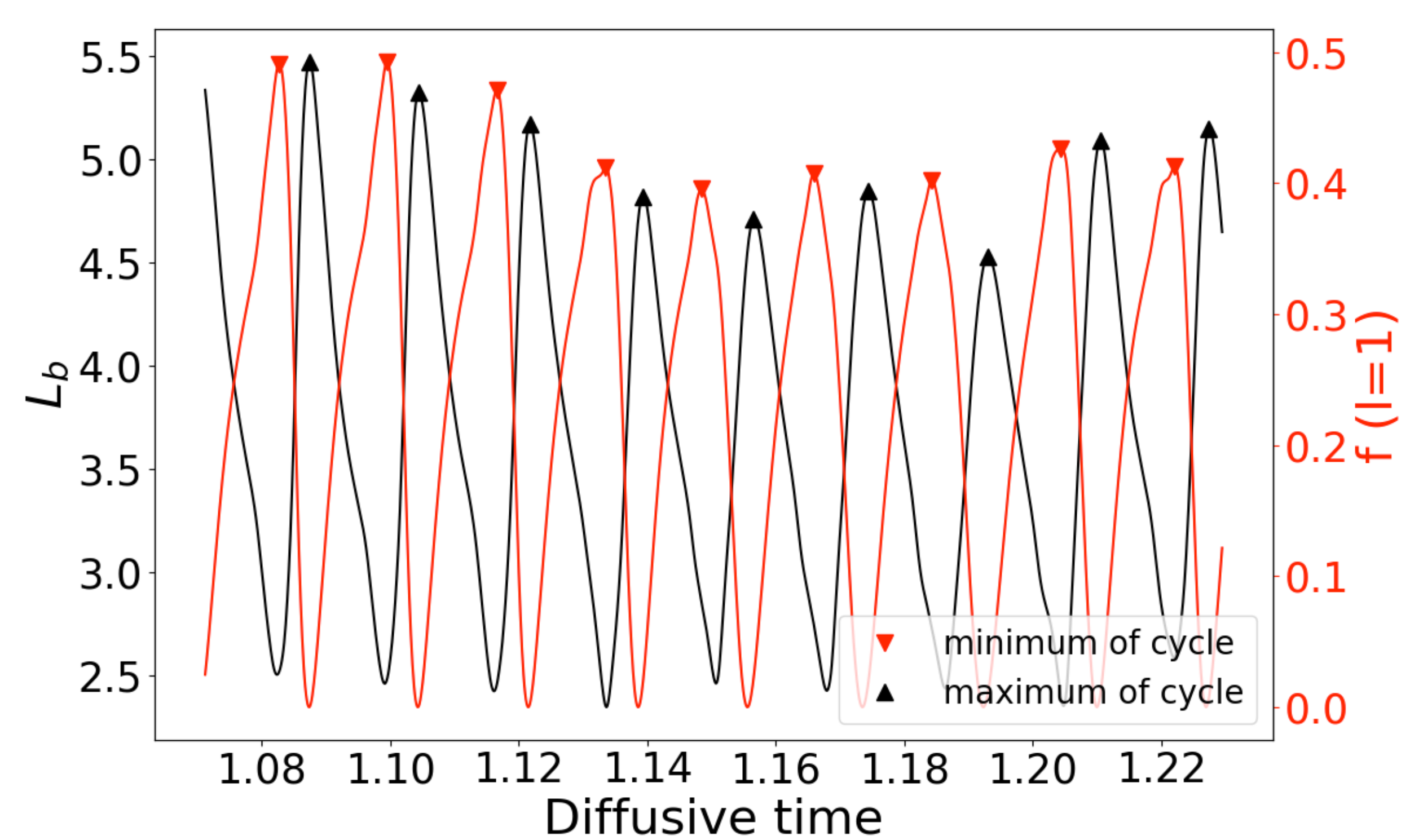}
    \caption{Time evolution of the magnetic characteristic length scale $l_b$ (in black) and the dipole coefficient $f_1$ (in red) for case DW2. Minima of activity are marked using red downward triangles, derived from $f_1$, and maxima with black upward triangles, derived from $l_b$.}
    \label{fig:lb_fdip_2w}
\end{figure}

 Figure \ref{fig:lb_fdip_2w} shows the time evolution of $l_b$ and $f_1$ for case DW2 with the two-way coupling. In this case, $l_b$ reaches slightly higher values, going up to 5.5. This is an indication that the cycle is slightly more multipolar than in case DW1. We see that the two curves are still in anti-correlation over the considered time period, however the relative position of the minima and maxima is affected by the two-way coupling. We can still see a phase quadrature between them at the beginning when the cycle is more anti-symmetrical like in case DW1. However when we start to be more symmetrical (at 1.15 $t_\eta$, when $f_1$ reaches only 40\% at maximum instead of 50\%), the minima and maxima of activity derived begin to be in anti-correlation because $l_b$ is also affected by the two-way coupling and the feedback of the wind.

\begin{figure}
    \centering
    \includegraphics[width=\textwidth]{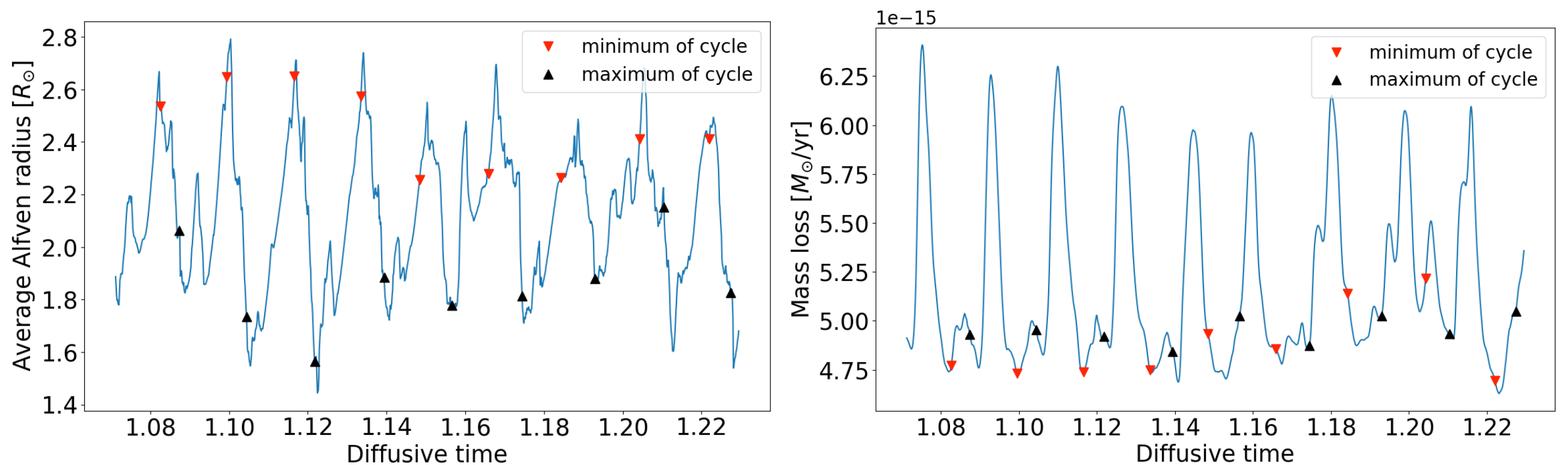}
    \caption{Time evolution of the integrated quantities of the wind for case DW2. From left to right: the average Alfvén radius $\langle r_A\rangle$ in $R_*$ and the mass loss $\dot{M}$ in $M_\odot/\rm{yr}$. Minima of activity are marked using red downward triangles, derived from $f_1$, and maxima with black upward triangles, derived from $l_b$.}
    \label{fig:ra_ml_2w}
\end{figure}

We can then see in figure \ref{fig:ra_ml_2w} the evolution of the integrated quantities defined above, namely the average Alfvén radius and the mass loss 
as the various cycles proceed. The Alfvén radius evolves between 1.5 and 2.75 $R_*$, which is a bit smaller as an interval with a smaller maximum value compared to case DW1. The mass loss evolves between 4.7 and 6.4 $10^{-15} \ M_\odot/\rm{yr}$, which is again a bit smaller with a higher minimum value compared to case DW1. The correlation with minima and maxima of activity is still working well when the cycle is anti-symmetrical, but when it tends to be more symmetrical (between 1.15 and 1.25 $t_\eta$) the trends are not so clear anymore. This could show that the trends derived in previous studies (described in section \ref{sec:evol}) are not generically true when the symmetry of the dynamo cycle changes, as $f_1$ stops to be the most relevant marker of the dynamics of the cycle.

\subsection{Impact of the interface region on dynamo modes}
\label{sec:modes}

As explained in the previous section, we see a clear difference between the two butterfly diagrams presented in figures \ref{fig:dp_snap_1w} and \ref{fig:dp_snap_2w}. Without the wind influence, the tachocline toroidal and surface radial magnetic fields are equatorially anti-symmetric and the cycle is regular with a period of $0.03t_\eta$. With the influence of the wind, the cycle has the same period but evolves to a more equatorially symmetric shape after 1.15 $t_\eta$ before returning to a more anti-symmetric pattern after 1.25 $t_\eta$. It however never reaches full anti-symmetry, which suggests a coupling between the dynamo families \citep{derosa_solar_2012}.

\begin{figure}
    \centering
    \includegraphics[width=\textwidth]{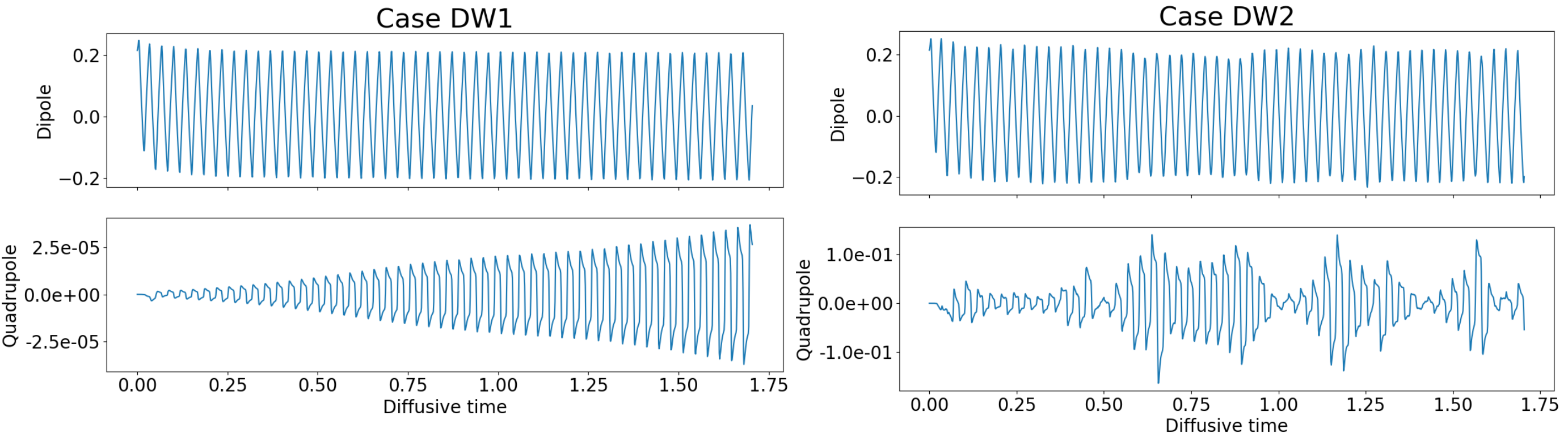}
    \caption{Evolution of the dipolar and quadrupolar modes for model DW1 (without wind feedback) and DW2 (with wind feedback) (respectively on the left and on the right). This is displayed in code units.}
    \label{fig:modes}
\end{figure}

To understand this difference, we show in Figure \ref{fig:modes} the evolution of the dipolar and quadrupolar modes ($\ell=1$ and $\ell=2$) for these two cases. The magnetic field is taken at the surface of the dynamo computational domain. We also show it for the entirety of the simulation, so between 0.0 and 1.75 $t_\eta$, while previous figures only focused on the time period between 1.0 and 1.3 $t_\eta$. Without the influence of the wind (left panel), the dipolar mode has a rather constant maximal amplitude around 0.2 code units with cyclic variations with a period of approximately 0.03 $t_\eta$ (same as the dynamo cycle),
while the quadrupolar mode has an amplitude 4 orders of magnitude smaller. Hence, although it is continuously growing, the symmetric family remains negligible. But with the influence of the wind in case DW2, the quadrupolar mode grows to an amplitude equivalent of the dipolar mode (0.1 versus 0.2). This is however not constant because we can clearly see a long-term modulation on the quadrupolar mode with periods of high intensity (between 1.15 and 1.25 $t_\eta$ as pointed before for example) and periods of very low values compared to the dipolar mode (between 1.0 and 1.15 $t_\eta$ for example). The feedback of the wind in this case has a visible influence by favoring the growth of the symmetric family by influencing the boundary conditions of the dynamo. This is a very interesting result, demonstrating the need to couple both ways the dynamo and the wind.

We nevertheless recall that the considered case has unusual values for a dynamo with $C_\alpha >> C_\Omega$. \cite{tavakol_structural_1995} have shown that in such parameter regimes, the dynamo is highly non-linear and can switch easily from symmetric to anti-symmetric regimes because of the quenching or an asymmetry which results in the coupling of the dynamo families. We have also performed a threshold study similar to \cite{jouve_role_2007} and have determined that for this set of parameters the quadrupolar mode has a growth threshold lower than the dipolar mode ($C_\alpha^Q=80$ versus $C_\alpha^D=100$). This case is therefore mostly a proof of concept to show that the feedback-loop between the dynamo and the wind plays an important role in our simulations. More investigations are needed to assess whether this feedback would be as efficient in other parameter regimes or for other types of dynamos.
Another possible explanation is the treatment of magnetic helicity in our interface which we will investigate in the next section.

\subsection{Analysis of helicity in coupled dynamo-wind simulations}

To quantify the exchange of information between the stellar interior and atmosphere, we now turn to magnetic helicity. Magnetic helicity is the measure of the warping and coiling of the magnetic field and is defined as:
\begin{equation}
    H = \int \mathbf{A}\cdot\mathbf{B}dV,
\end{equation}
where $\mathbf{B}$ is the magnetic field and $\mathbf{A}$ is the vector potential associated to $\mathbf{B}$ such as $\mathbf{B} = \nabla\times\mathbf{A}$. 
There is no unicity of the vector potential $\mathbf{A}$, which means that magnetic helicity has to be computed with respect to a certain gauge. To avoid this problem we can define the relative magnetic helicity which is gauge-invariant \citep{berger_introduction_1999}:
\begin{equation}
    H_{rel} = \int \left(\mathbf{A} + \mathbf{A}_p\right)\cdot\left(\mathbf{B} - \mathbf{B}_p\right)dV,
\end{equation}
where $\mathbf{B}_p$ is a reference component, chosen to be potential ($\nabla\times\mathbf{B}_p=0$) and which components match the components of $\mathbf{B}$ at the boundaries of the domain
($\mathbf{B}_p\cdot\mathbf{n}|_S = \mathbf{B}\cdot\mathbf{n}|_S$). $\mathbf{A}_p$ is the vector potential associated to $\mathbf{B}_p$ such as $\mathbf{B}_p = \nabla\times\mathbf{A}_p$.

To compute $\mathbf{B}_p$ and $\mathbf{A}_p$ we use the fact that $\mathbf{B}_p$ is chosen to be potential, which implies that it is deriving from a scalar magnetic potential $\phi$ ($\mathbf{B}_p = \Delta\phi$). $\phi$ can be projected onto spherical harmonics and the boundary conditions then give us enough constraints to determine it. To avoid discontinuity problems at the interface we compute a reference field for the stellar interior and another for the corona, and we have checked after computation that the two match. To compute $\mathbf{A}$, we combine direct integration of the magnetic field and a projection onto spherical harmonics to solve equations on the current $\mathbf{j}$. For more details, please refer to appendix \ref{appendix:hel}.

\begin{figure}
    \centering
    \includegraphics[width=\textwidth]{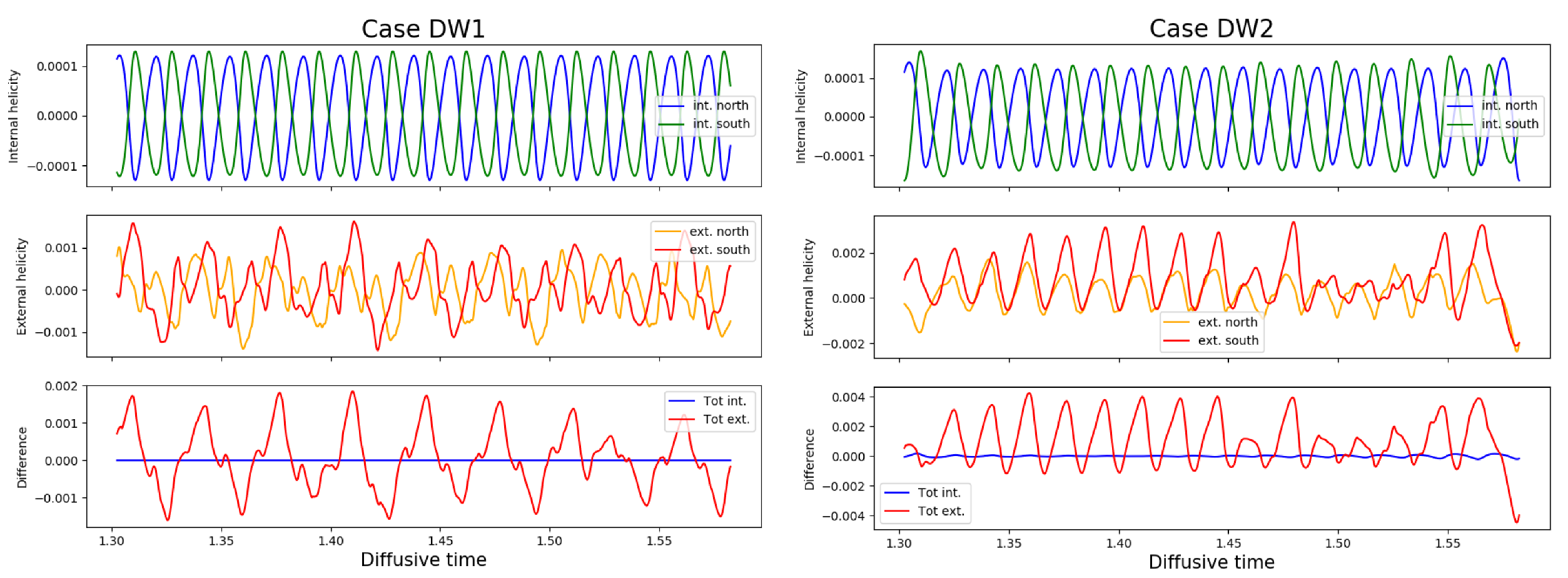}
    \caption{Time evolution of relative magnetic helicity for case DW1 (on the left) and DW2 (on the right), in code units. The top panel shows the helicity inside the star, and for the northern and southern hemisphere (in blue and green, respectively). The middle panel shows the same separation but for the helicity outside the star (northern hemisphere in orange, southern hemisphere in red). The last panel shows the difference between the northern and southern internal relative helicity in blue, and the difference between the northern and southern external relative helicity in red.}
    \label{fig:hel_rel}
\end{figure}

Figure \ref{fig:hel_rel} shows the time evolution of the relative magnetic helicity for case DW1 (on the left) and case DW2 (on the right). We have selected a different time interval as what was previously shown, this time from $1.3$ to $1.58 t_\eta$. This is to avoid the region where the dynamo modes are too coupled to really study only the influence of the interface on magnetic helicity in the most basic situation. We have divided it into stellar interior (in blue/green colors) and exterior (in red/orange colors).
We also decompose the helicity between its northern and southern components, respectively in blue and green for the interior and in orange and red for the exterior. This allows us to check the most basic property of magnetic helicity, which is its conservation in the numerical domain when magnetic diffusion is neglected. In the figure, we can see that indeed the northern and southern components of helicity apparently compensate each other relatively well at each time for the star interior, and this for both coupling cases. This shows that the diffusivity inside the star is not dominant enough to suppress this property on the time-scales considered here. We also can see that the helicity is slightly more dominant in the exterior of the star with one order of magnitude of difference with the interior. However in the stellar corona, the conservation of relative helicity is visibly not as good as inside the star.
We can see in the third panel, which shows the amplitude of the difference between northern and southern hemisphere relative helicity for the stellar interior (in blue) and the exterior (in red) that the conservation of helicity in the exterior is not achieved in DW1 nor DW2. Ohmic diffusion does affect the relative helicity inside the star, but is unlikely to affect it in the corona where it weakly acts only in the very first points of the wind domain. This points to a possible loss or injection of helicity at the interface and/or outer boundary. We will then analyze the fluxes of helicity to check this assumption. 

Our set-up was designed to let only the poloidal field go through the dynamo-wind interface. Now we want to see how our boundary condition affects magnetic helicity transfer into the idealized corona, as it has been advocated for instance by \cite{low_magnetic_2013}. To do so, we compute the evolution equation for $H_{rel}$ \citep{barnes_mechanical_1988}:
\begin{equation}
    \frac{\partial H_{rel}}{\partial t} = -\frac{8\pi \eta}{c} \int \mathbf{j}\cdot\mathbf{B}dV + 2\int_{\partial V} \left[\mathbf{A}_p\times\left(-\mathbf{u}\times\mathbf{B}+\eta\nabla\times\mathbf{B}\right)\right]\cdot \mathbf{n}dS = -\frac{8\pi \eta}{c} \int \mathbf{j}\cdot\mathbf{B}dV + 2F_H,
    \label{eq:dhdt}
\end{equation}
where $\mathbf{j}$ is the current associated to the magnetic field $\mathbf{B}$.

Based on this equation, we can define a helicity flux through a surface associated to a temporal variation of helicity. Under the assumptions of axisymmetry, 2.5D, and spherical coordinates of this study, this yields:
\begin{equation}
    F_H = 2\pi \int_0^\pi \left(\left(u_\phi B_r - u_rB_\phi\right) - \eta (\nabla\times\mathbf{B})_\theta\right)A_{p,\phi}R_s^2\rm{sin}\ \theta d\theta,
\end{equation}
where $R_s$ is the radius of the spherical surface through which the flux is computed. For more details about helicity budget, please refer to appendix \ref{appendix:hel}. 

\begin{figure}
    \centering
    \includegraphics[width=\textwidth]{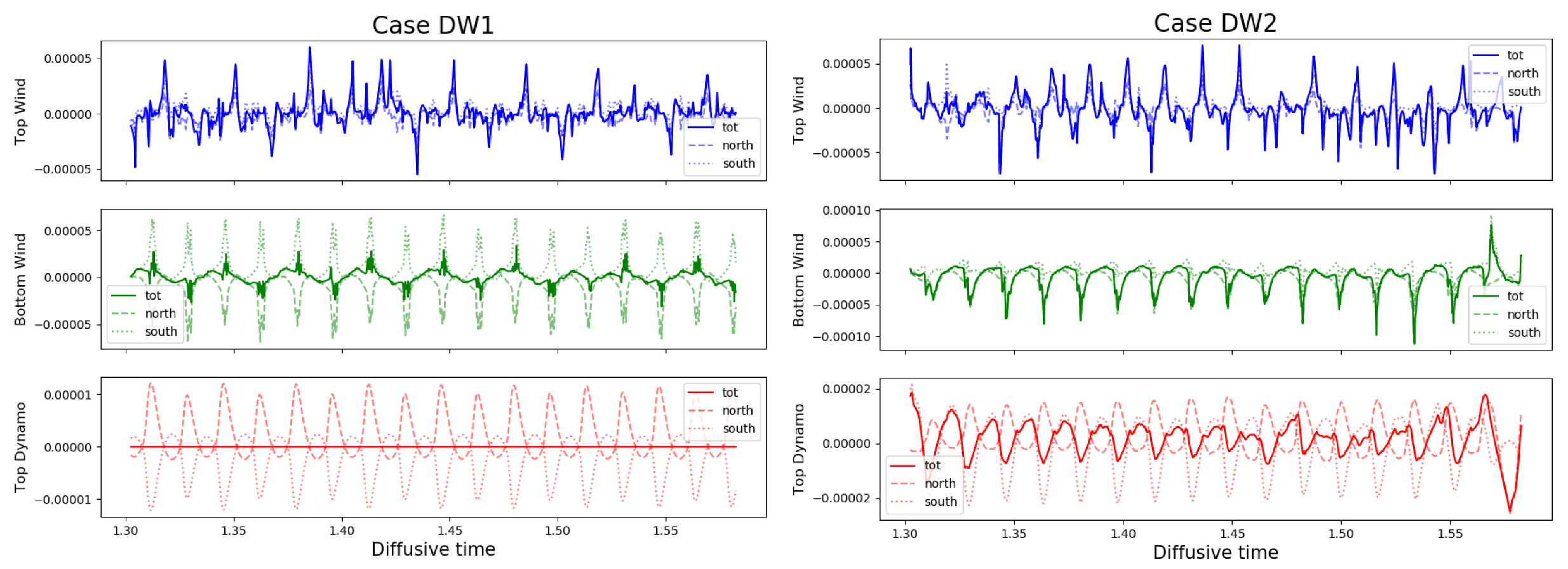}
    \caption{Relative magnetic helicity fluxes for case DW1 (on the left) and DW2 (on the right) in code units. They are shown at three different radii: top of the wind computational domain ($r=20R_*$) in blue, bottom of the wind domain (just above the interface at $r=R_*$) in green and top of the dynamo domain (just below the interface at $r=R_*$) in red. The top and middle panels show the fluxes out of the wind domain (a positive/negative value means helicity is decreasing/increasing in the wind domain). The bottom panel shows the flux out of the dynamo top boundary (a positive/negative value means helicity is decreasing/increasing in the dynamo domain). Note how the magnetic helicity fluxes at the top dynamo boundary (bottom panel) differs between DW1 and DW2 models. In DW1 there is no net transfer, while there is a clear inward flux in DW2.}
    \label{fig:hel_flux}
\end{figure}

Figure \ref{fig:hel_flux} shows the time evolution of $F_H$ at the three different radii for case DW1 (on the left) and DW2 (on the right): in blue we have the top boundary condition for the wind at $r=20R_*$; in green we have the bottom of the wind domain (just above the interface at $r=R_*$); finally, in red, we have the top of the dynamo domain (just below the interface at $r=R_*$). The two upper panel show fluxes for the helicity budget in the wind domain: a positive (negative) value means that the wind helicity decreases. The bottom panel show the flux at the top of the dynamo domain: a positive (negative) value means that the dynamo helicity decreases. In all cases we show the decomposition of the helicity fluxes on the northern (in dashed) and southern (in dots) hemisphere. As a sanity check, we have evaluated the flux loss at the bottom boundary condition of the dynamo, and found it to be completely negligible. At the top wind boundary condition, helicity is lost due to the modified outflow boundary condition for the wind. We find a loss of the order of $5 \ 10^{-5}$ (in code units) at maximum in both cases. With these numbers in mind, we can then see that for case DW1, the injection at the interface is negligible, being practically zero at the top of the dynamo and reaching only $10^{-5}$ at maximum at the bottom of the wind domain. However for case DW2, helicity transfer is enhanced at the interface with typical amplitudes of $10^{-4}$ at the bottom of the wind domain. We also notice that the northern and southern fluxes are different from one another in case DW2 due to the asymmetry discussed before (see section \ref{sec:modes}), while they compensate perfectly in case DW1. Note that in both cases we have verified that ohmic dissipation plays a negligible role, as shown in Appendix \ref{appendix:hel}.

We therefore see that our interface plays a non-trivial role in helicity conservation and injection. The net helicity flux at the top of the dynamo domain is small in case DW1, but a substantial helicity is injected at the bottom of the wind domain due to our boundary condition on $B_\phi$. In case DW2, this effect is even more pronounced as our boundary dictates the helicity budget in the wind domain.  
It furthermore leads to helicity injection in the dynamo domain, which adapts strongly in case DW2 with the two-way coupling. This is probably another factor which explains why the coupling of the modes is so strong in this case. This means that we need to find a more appropriate way to limit the generation of currents at the bottom of our wind domain without jeopardizing the conservation of helicity.
We plan to thus investigate further this condition in the interface for a more solar parameter space where the physical quantities may be less sensitive to this effect.

\section{Conclusion and discussion}
\label{sec:conclusion}

We have presented here one of the first example of self-consistent numerical coupling between an $\alpha$-$\Omega$ stellar dynamo and a thermally driven stellar wind. We used the PLUTO code \citep{mignone_pluto:_2007} to couple the wind set-up of \cite{reville_solar_2015} adapted for spherical coordinates in \cite{perri_simulations_2018}, and the dynamo model described in the benchmark of \cite{jouve_solar_2008} and implemented for the first time in the PLUTO code. The key-point of this model is the interface between the two physical domains: instead of being a software coupling between two independent codes, we have used a single numerical domain where a four-mesh thick layer structure at the surface of the star has been implemented. This allows us to solve self-consistently for the dynamo-wind coupling inside a computational domain encompassing the star's interior and atmosphere from $0.6R_*$ up to $20R_*$. We can thus study dynamically the physical exchanges between the dynamo and the wind along a dynamo cycle. The magnetic field generated inside the star is extrapolated using a potential field extrapolation across the interface layers to be transmitted to the corona and affect the structure of the wind by its changes of amplitude and topology. We present here one of the first models able to quantify the feedback loop between the wind and the internal magnetic field.

Here we lay out only the first developments of this model which must be seen as a proof of concept to demonstrate the efficiency and interest of this method. To do so, we presented a model with an activity cycle having a much shorter period than the Sun.
This model demonstrates cyclic variations of the corona along the dynamo cycle. These variations are associated with the opening and closing of coronal holes with the corresponding generation of streamers and variations of the wind integrated quantities such as the average Alfvén radius or the mass loss. These correlate with the minima and maxima of activity of the cycles, reproducing results obtained so far only by data-driven models \citep{merkin_time-dependent_2016, reville_global_2017} or quasi-static approaches \citep{pinto_coupling_2011, perri_simulations_2018}. This approach however may generate transients from which it is difficult to evaluate if they carry physical information. By adjusting the interface layer, we are able to enable a one-way or two-way coupling between the dynamo and the wind, and thus we are able to illustrate a possible feedback loop between the dynamo and the wind. By enabling the feedback of the wind, we observe a different dynamo behaviour where the wind feedback loop seems to enhance the secondary family modes
\citep{ossendrijver_solar_2003, svalgaard_asymmetric_2013}. We also demonstrate that our interface region may generate additional helicity due to the various conditions on $B_\phi$, and are exploring new methods to modify this property. By focusing next on slower dynamo cycles we may have less transients and reconnection, which could also lead to more stable models.

There are many different directions to broaden the accuracy and the applications of this model. To have a better description of the physics, we have begun to implement a Babcock-Leighton mechanism with flux transport for the dynamo: this means that instead of the deep-seated $\alpha$ effect, we take into account the generation of strong toroidal magnetic tubes at the base of the convective zone and their rise in this model by buoyancy, then twist by the Coriolis force, and then allow the regeneration of poloidal field at the surface. The meridional circulation redistributes the poloidal field at the base of the convective zone to be sheared by the differential rotation \citep{babcock_topology_1961, leighton_magneto-kinematic_1969, dikpati_babcock-leighton_1999, jouve_role_2007}. This model needs the implementation of a non-local term which is more complex given the architecture of the PLUTO code; we will report in a future work this new dynamo set-up. The description of the wind can also be improved by using a more realistic heating: instead of a polytropic equation, there are models based on turbulence and Alfvén-wave heating that give results in good agreement with observations \citep{riley_inferring_2015, reville_role_2020, hazra_modeling_2021}. For our next study, we will add these new elements and tackle the description of a coupled model between a solar-like dynamo with an 11-year cycle and a solar-like turbulent corona, with an interface better designed to avoid injection of helicity. This is of course a numerical challenge given the discrepancy of the timescales with a timestep of the order of the hour to simulate at least 11 years of solar activity. On the other hand it will allow easier comparison with observational data to better calibrate the model \citep{mccomas_weaker_2008}. The parameter space description given in Appendix \ref{appendix:osc} will help us ensure that the coupling is operating in an efficient and physical way. One of the more ambitious objectives would be also to describe more precisely the transition region between the photosphere and the corona. In our model, the resolution is too coarse to allow a separate description of the photosphere, chromosphere and corona, and the transitions in density and temperature between these regions \citep{gary_plasma_2001, meyer-vernet_basics_2007}. 
Finally we remind the reader that this model could actually be very easily used for other stars of F, G or K spectral type with an internal structure similar of that of the Sun (with a convective zone at the surface and a radiative zone deep inside) and a thermal-driven wind. With elements such as the rotation profile, the internal poloidal flows or the dynamo mechanism prescribed with input parameters, we can easily change them to reproduce other stars. For example the two cases described in this article were chosen purely theoretically but can be assimilated to young stars of T-Tauri type, with results similar to the work of \cite{von_rekowski_stellar_2006}. \\ 

\textit{Acknowledgments}\\

We thank Patrick Hennebelle and Thierry Foglizzo for useful suggestions. This work was supported by a CEA 'Thèse Phare' grant, by CNRS and INSU/PNST program, by CNES Solar Orbiter and Météo de l'espace funds and by the ERC Synergy grant WholeSun \#810218. Computations were carried out using CEA CCRT and CNRS IDRIS facilities within the GENCI 40410133, 60410133, and 80810133 allocations, and a local meso-computer founded by DIM ACAV+.

\appendix

\section{Normalization of the equations}
\label{appendix:norm}

Dynamo and wind numerical codes usually have different normalisations for the MHD equations, leading to different control parameters for simulations. For the coupling, as dynamo and wind happen in the same numerical domain, we had to choose one normalisation and convert the corresponding parameters. We chose the wind normalisation to focus only on the induction equation in the PLUTO code. In this section, the two different normalisations are presented along with the conversion factors between the two for the induction equation.

We start with the induction equation from the mean field theory \citep{moffatt_magnetic_1978}:
\begin{equation}
\frac{\partial\mathbf{B}}{\partial t} = \nabla\times\left(\mathbf{U}_{pol}\times\mathbf{B} + \mathbf{U}_\phi\times\mathbf{B} - \eta\nabla\times\mathbf{B}+\alpha\mathbf{B}\right).
\label{eq:induction}
\end{equation}

Each quantity can then be written as a product between a constant physical amplitude with units and a normalized profile describing the dependency on $(r,\theta,\phi,t)$:
\begin{equation}
t = t_0\tilde{t}, r = L_0\tilde{r}, \mathbf{B} = B_0\widetilde{\mathbf{B}}, \mathbf{U}_{pol} = U_0\widetilde{\mathbf{U}}, \mathbf{U}_{\phi} = \Omega_0L_0\widetilde{\mathbf{U}}_\phi, \eta = \eta_0\tilde{\eta}, \alpha = \alpha_0\tilde{\alpha}.
\label{eq:norm}
\end{equation}

By injecting the decomposition described in \ref{eq:norm} in equation \ref{eq:induction}, we obtain the following general normalised induction equation:
\begin{equation}
\frac{\partial\widetilde{\mathbf{B}}}{\partial \tilde{t}} = \nabla\times\left( \frac{t_0U_0}{L_0} \widetilde{\mathbf{U}}_{pol}\times\widetilde{\mathbf{B}} + t_0\Omega_0 \widetilde{\mathbf{U}}_\phi\times\widetilde{\mathbf{B}} - \frac{t_0\eta_0}{L_0^2} \tilde{\eta}\nabla\times\widetilde{\mathbf{B}} + \frac{\alpha_0t_0}{L_0} \tilde{\alpha}\widetilde{\mathbf{B}}\right).
\label{eq:induction_norm}
\end{equation}

To obtain the usual dynamo normalisation, we set the time and length constants to respectively the diffusive time and the solar radius:
\begin{equation}
t_0 = \frac{R_*^2}{\eta_t}, L_0 = R_*,
\label{eq:norm_dynamo}
\end{equation}
with $R_* = 6.96 \ 10^{10} \ \rm{cm}$ and $\eta_t = 10^{11} \ \rm{cm}^{2}\rm{s}^{-1}$. By injecting \ref{eq:norm_dynamo} into equation \ref{eq:induction_norm}, this leads to the following standard form of the induction equation:
\begin{eqnarray}
\frac{\partial\widetilde{\mathbf{B}}}{\partial \tilde{t}} &=& \nabla\times\left( \frac{U_0R_*}{\eta_t} \widetilde{\mathbf{U}}_{pol}\times\widetilde{\mathbf{B}} + \frac{\Omega_0R_*^2}{\eta_t} \widetilde{\mathbf{U}}_\phi\times\widetilde{\mathbf{B}} - \frac{\eta_0}{\eta_t} \tilde{\eta}\nabla\times\widetilde{\mathbf{B}} + \frac{\alpha_0R_*}{\eta_t} \tilde{\alpha}\widetilde{\mathbf{B}}\right) \nonumber \\
&=& \nabla\times\left( R_e \widetilde{\mathbf{U}}_{pol}\times\widetilde{\mathbf{B}} + C_\Omega \widetilde{\mathbf{U}}_\phi\times\widetilde{\mathbf{B}} - \frac{\eta_0}{\eta_t} \tilde{\eta}\nabla\times\widetilde{\mathbf{B}} + C_\alpha \tilde{\alpha}\widetilde{\mathbf{B}}\right),
\label{eq:induction_dynamo}
\end{eqnarray} 
with the standard control parameters being the magnetic Reynolds number $R_e$, the normalized rotation rate $C_\Omega$ and the normalized alpha effect amplitude $C_\alpha$. 

For the wind normalisation used in the PLUTO code, we usually set the length, speed and time constants to respectively the solar radius, the Kepler speed at the surface of the star and the time ratio obtained by these two values:
\begin{equation}
L_0 = R_*, U_0 = U_K = \sqrt{\frac{GM_*}{R_*}}, t_0 = \frac{L_0}{U_0},
\label{eq:norm_wind}
\end{equation}
with $G = 6.67 \ 10^{-8} \ \rm{dyne}.\rm{cm}^2.\rm{g}^{-2}$ and $M_*=M_\odot=1.99 \ 10^{33} \ \rm{g}$. By injecting \ref{eq:norm_wind} into equation \ref{eq:induction_norm}, we then obtain a new normalisation for the induction equation:
\begin{eqnarray}
\frac{\partial\widetilde{\mathbf{B}}}{\partial \tilde{t}} &=& \nabla\times\left( \frac{U_0}{U_K} \widetilde{\mathbf{U}}_{pol}\times\widetilde{\mathbf{B}} + \frac{\Omega_0R_*}{U_K} \widetilde{\mathbf{U}}_\phi\times\widetilde{\mathbf{B}} - \frac{\eta_0}{U_KR_*} \tilde{\eta}\nabla\times\widetilde{\mathbf{B}} + \frac{\alpha_0}{U_K} \tilde{\alpha}\widetilde{\mathbf{B}}\right) \nonumber \\
&=& \nabla\times\left( R_e^P \widetilde{\mathbf{U}}_{pol}\times\widetilde{\mathbf{B}} + C_\Omega^P \widetilde{\mathbf{U}}_\phi\times\widetilde{\mathbf{B}} - \eta^P \tilde{\eta}\nabla\times\widetilde{\mathbf{B}} + C_\alpha^P \tilde{\alpha}\widetilde{\mathbf{B}}\right).
\label{eq:induction_wind}
\end{eqnarray}

To switch from dynamo to wind normalisation (\textit{ie} from equation \ref{eq:induction_dynamo} to equation \ref{eq:induction_wind}), the following conversion factor can be applied:
\begin{equation}
\frac{\partial\widetilde{\mathbf{B}}}{\partial \tilde{t}} = \frac{\eta_t}{R_* U_K}\nabla\times\left( R_e \widetilde{\mathbf{U}}_{pol}\times\widetilde{\mathbf{B}} + C_\Omega \widetilde{\mathbf{U}}_\phi\times\widetilde{\mathbf{B}} - \frac{\eta_0}{\eta_t} \tilde{\eta}\nabla\times\widetilde{\mathbf{B}} + C_\alpha \tilde{\alpha}\widetilde{\mathbf{B}}\right).
\label{eq:induction_passage}
\end{equation}

\section{Extrapolation of the poloidal magnetic field}
\label{appendix:extra}

In our interface region between the dynamo and the wind computational domains, we need to extrapolate the magnetic field generated by the dynamo inside the star to quantify the impact of its amplitude and topology on the wind. To do so, we have implemented a potential field extrapolation using the following mathematical method.

We choose to perform a potential field extrapolation, meaning that the field at the surface of the star derives from a scalar potential $\Phi$ such as $\mathbf{B} = -\nabla\Phi$ which corresponds to a magnetic field in vacuum. Such a magnetic field has the property that both its divergence and curl are equal to 0, thus $\Phi$ is a solution to the Laplace equation which means it can be decomposed as a sum of spherical harmonics $Y_\ell^m$:
\begin{equation}
\Phi(r,\theta,\phi) = \sum_{\ell=0}^{\ell_{max}}\sum_{m=0}^\ell \left(\Phi_{\ell,m}^ar^\ell + \Phi_{\ell,m}^b r^{-(\ell+1)}\right)Y_\ell^m(\theta,\phi),
\end{equation}
where $Y_\ell^m(\theta,\phi) = c_\ell^mP_\ell^m(\theta)e^{im\phi}$, $c_\ell^m = \sqrt{\frac{2\ell+1}{4\pi}\frac{(\ell-m)!}{(\ell+m)!}}$ and $P_\ell^m(\theta,\phi)$ are the associated Legendre polynomials. As the first branch in $r^\ell$ is divergent, we keep only the second one, choosing this form for the scalar potential:
\begin{equation}
    \Phi(r,\theta,\phi) = \sum_{\ell=0}^{\ell_{max}}\sum_{m=0}^\ell \Phi_{\ell,m}^b r^{-(\ell+1)}Y_\ell^m(\theta,\phi).
\end{equation}

Applying the gradient operator to obtain the magnetic field and projecting on the vectorial spherical harmonics $\mathbf{R}_\ell^m = Y_\ell^m(\theta,\phi)\mathbf{e}_r$ and $\mathbf{S}_\ell^m = \nabla_{\perp}Y_\ell^m(\theta,\phi) = \frac{\partial Y_\ell^m}{\partial\theta}(\theta,\phi)\mathbf{e}_\theta + \frac{1}{\rm{sin}\theta}\frac{\partial Y_\ell^m}{\partial\phi}(\theta,\phi)\mathbf{e}_\phi$ yields:
\begin{eqnarray}
\mathbf{B}(r,\theta,\phi) &=& \sum_{\ell=0}^{\ell_{max}}\sum_{m=0}^\ell  \Phi_{\ell,m}^b(\ell+1)r^{-(\ell+2)}\mathbf{R}_\ell^m(\theta,\phi) - \Phi_{\ell,m}^b r^{-(\ell+2)}\mathbf{S}_\ell^m(\theta,\phi), \nonumber \\
&=& \sum_{\ell=0}^{+\infty}\sum_{m=0}^{\ell} \alpha_{\ell,m}(r)\mathbf{R}_\ell^m(\theta,\phi) + \frac{\beta_{\ell,m}(r)}{\ell+1}\mathbf{S}_\ell^m(\theta,\phi).
\label{eq:b_vec_spharm}
\end{eqnarray}

We then assume that we know the value of the magnetic field at a certain radius named $R_b$:
\begin{equation}
\mathbf{B}(r=R_b) = \sum_{\ell=0}^{+\infty}\sum_{m=0}^\ell \alpha_{\ell,m}(R_b)\mathbf{R}_\ell^m + \frac{\beta_{\ell,m}(R_b)}{\ell+1}\mathbf{S}_\ell^m.
\end{equation}

This yields a system of two equations but with only one variable that is $\Phi^b_{\ell,m}$. We have to choose to base ourselves on either $B_r$ or $B_\theta$ to compute the solution. Based on previous benchmarks like \cite{jouve_solar_2008} we realized that most dynamo models use $B_r$ as a reference, so that is what we chose as well. This yields then the solution:
\begin{equation}
    \Phi^b_{\ell,m} = \frac{\alpha_{\ell,m}(R_b)}{\ell+1}R_b^{\ell+2}.
\label{eq:phia_phib}
\end{equation}

By injecting \ref{eq:phia_phib} into equation \ref{eq:b_vec_spharm}, this yields in the end:
\begin{eqnarray}
\alpha_{\ell,m}(r>R_b) &=& \alpha_{\ell,m}(R_b)\left(\frac{r}{R_b}\right)^{-(\ell+2)}, \\
\beta_{\ell,m}(r>R_b) &=& - \alpha_{\ell,m}(R_b)\left(\frac{r}{R_b}\right)^{-(\ell+2)}.
\end{eqnarray}

\section{Preliminary study with an oscillatory dipole}
\label{appendix:osc}

Before diving into the full coupling with a dynamo cycle, we started with a preliminary study where we did not use yet the coupling model described above. The goal of this section is to have a more precise idea of the parameter space allowed for our model for the coupling to be operational. This will give a first indirect characterization of our coupling model by discussing the issue of time scales.

\begin{figure}
    \centering
    \includegraphics[width=\textwidth]{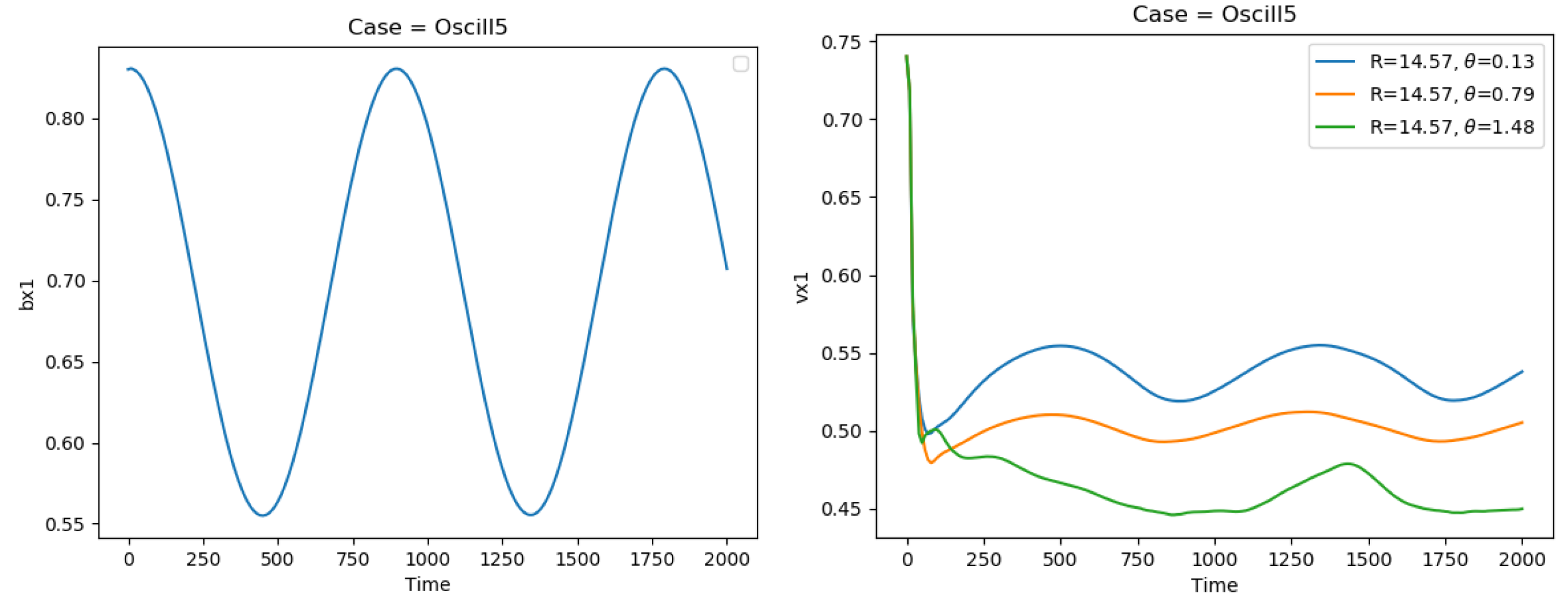}
    \caption{Time evolution in 1D of our oscillatory dipole. The left panel shows the evolution of $B_r$ which is imposed at the surface of the star. The right panel shows the response in the wind speed $u_r$ at three different latitudes (near the poles, at mid-latitudes and near the equator).}
    \label{fig:oscill_1d}
\end{figure}

In this simplified model, we have only the wind computational domain with the boundary conditions described in figure \ref{fig:wind_pic_bc}. The magnetic field is imposed at the surface of the star, but it varies in time: at each timestep, it oscillates at a certain frequency $\omega$ which is a control parameter. The amplitude of the field is then given in code units by the following relation:
\begin{equation}
B_{osc} = \frac{u_A}{u_{esc}}\sqrt{8\pi\rho_*}\left(1 + 0.2\cos\left(t\omega\right)\right),
\end{equation}
where $u_A/u_{esc}$ is the Alfvén speed at the equator and at the surface of the star divided by the escape velocity, $\rho_*$ is the density at the surface of the star. We recall that the Alfvén speed $u_A$ is given by $|\mathbf{B}|/\sqrt{4\pi\rho}$. That way the magnetic field never reaches a null amplitude, it oscillates between 80\% and 120\% of its initial value. These oscillations are shown in the left panel of Figure \ref{fig:oscill_1d}. The wind then adapts to the magnetic field configuration: the right panel of Figure \ref{fig:oscill_1d} shows the radial speed $u_r$ of the wind at $r=14.57 R_*$ (so at 75\% of our numerical box) and at three different latitudes ($\theta=7\degree$ which is near the pole, $\theta=45\degree$ which is at mid-latitude and $\theta=84\degree$ which is near the equator). We can see oscillations with a frequency corresponding to the $\omega$ frequency of the magnetic field given in the left panel. Since we studied only a dipolar topology for the magnetic field, we have a latitudinal dependency, which means that the oscillations are more easily transmitted near the poles where the magnetic field influence is weaker so that the wind can adapt quicker, while the oscillations are less visible at the equator where the dipole is strong and thus creates a dead zone where the wind speed is reduced (between 0.45 and 0.48 vs 0.50 and 0.55 in PLUTO units). 

\begin{figure}
    \centering
    \includegraphics[width=\textwidth]{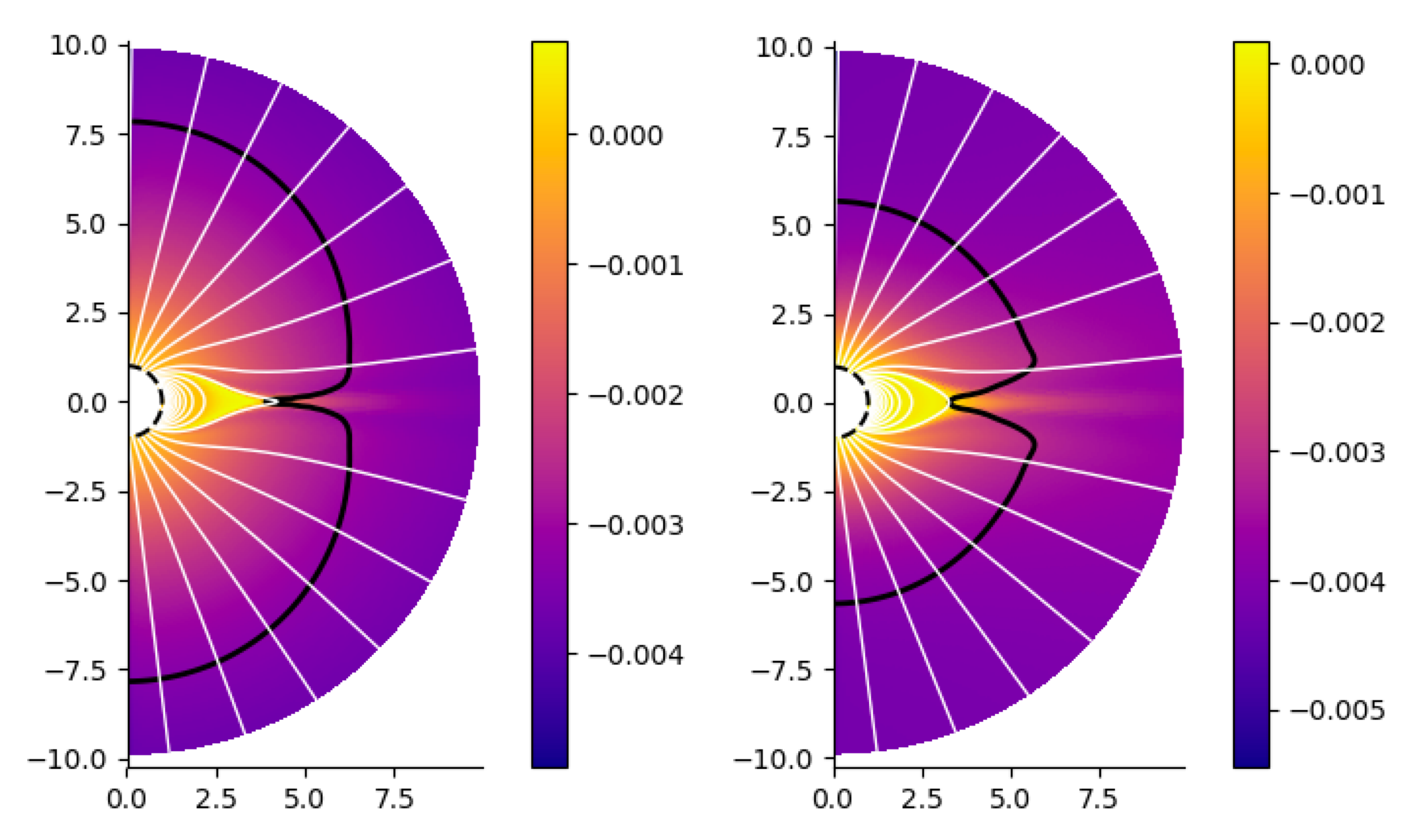}
    \caption{Meridional cuts in 2D of the oscillatory dipole model. The left panel shows the case when the dipole amplitude is minimal (at 80\% of its initial value), and the right panel the case when the dipole amplitude is maximal (at 120\% of its initial value). The color bar represents the quantity $\mathbf{u}\cdot\mathbf{B}/(c_s||\mathbf{B}||)$, which is the wind speed projected on the magnetic field normalized by the Mach number. The white lines represent the poloidal magnetic field lines. The black line represents the Alfvén surface which is the surface at which $u_r=u_A$. The star surface is represented by a black dashed line.}
    \label{fig:oscill_2d}
\end{figure}

To quantify the propagation of information, we introduce three characteristic times: 
\begin{equation}
t_A = \frac{\Delta r}{u_A}, t_{cyc} = \frac{2\pi}{\omega}, t_\eta = \frac{(\Delta r)^2}{\eta},
\label{eq:oscill_times}
\end{equation} 
where $t_A$ is the local Alfvén time, $t_{cyc}$ is the time associated with the oscillation of the surface magnetic field and $t_\eta$ is the magnetic diffusive time associated with $\eta$ in the wind computational domain. Without diffusivity, our various simulations yield the following empirical criterion:
\begin{equation}
t_{A} < 0.01 t_{cyc}.
\label{eq:osc_crit1}
\end{equation}
This is a first criterion concerning the resolution used for the numerical grid (cf. equation \ref{eq:oscill_times}). For the tests performed, we fixed the resolution with the following grid: we used an uniform grid of 256 points in $\theta$ between 0 and $\pi$, and a stretch grid of 256 points in $r$ between 1 and 20 $R_*$ with $\Delta r=0.01$ at the surface of the star. We then used various Alfvén speeds to vary the corresponding Alfvén time. Figure \ref{fig:oscill_2d} shows how the wind is supposed to react in a case where the criterion \ref{eq:osc_crit1} is respected and thus the coupling is operating properly. In the two meridional cuts, we can clearly see how the wind has adapted itself to the dipolar magnetic field: we have an equatorial streamer up to 4 $R_*$, elsewhere the magnetic field lines (in white) are open \footnote{Please note that this is a misuse of language : the Maxwell laws prohibit the existence of a magnetic monopole, meaning that the magnetic field lines are not really "open", they simply reconnect far away from the star and way outside our numerical box, thus only appearing open in our simulations.}. The black line is the Alfvén surface, which is the surface at which $u_r=u_A$. This is the major visible change between the two configurations: on the left, the amplitude of the magnetic field is minimal (80\% of its initial value), thus the Alfvén surface is further away from the star (on latitudinal average, the corresponding Alfvén radius $r_A$ is equal to 5.0 $R_\star$), while on the right the amplitude of the magnetic field is maximal (120\% of its initial value) making the Alfvén surface closer to the star ($r_A=3.5 R_\star$). 

If we choose to add magnetic diffusivity to the wind model, it can help propagate the information from inside the star to the outside. If the diffusivity is limited to a few grid points above the star surface (like with the profile described in equation \ref{eq:eta_profile}), then it has little to no influence to the corresponding relaxed state of the wind. We still have to respect another criterion to make sure that the wind has time to adapt to the magnetic cycle:
\begin{equation}
t_\eta < 0.01 t_{cyc}.
\label{eq:osc_crit2}
\end{equation} 

In the cases which did not respect criterion \ref{eq:osc_crit1} and criterion \ref{eq:osc_crit2}, the coupling was not working properly. What we observed was that the magnetic field was evolving too quickly at the star surface so that the wind could not adapt fast enough. This led to an uncoupling and thus a drop in the magnetic field, especially in the latitudinal component of the magnetic field $B_\theta$. This leads to a strong back-reaction of the Lorentz force which is expressed as:
\begin{equation}
    \mathbf{F}_L = \frac{1}{4\pi}\nabla\times\mathbf{B}\times\mathbf{B} \\
    \Rightarrow 4\pi F_{L,\theta} = \left(\frac{1}{r}\frac{\partial}{\partial r}\left(rB_\theta\right) - \frac{1}{r}\frac{\partial B_r}{\partial\theta}\right)B_\theta - \frac{1}{r\rm{sin}\theta}\frac{\partial}{\partial\theta}\left(\rm{sin}\theta B_\phi\right)B_r.
\end{equation}
This force triggers latitudinal flows, and in the end the wind speed is compressed to negative flows because of it, which leads to accretion at the poles.

\section{Computations for helicity}
\label{appendix:hel}

\subsection{Computation of $A_p$ with various boundary conditions}

\subsubsection{General principle}

The relative magnetic helicity, which is gauge invariant, is defined as:
\begin{equation}
H_{rel} = \int (\mathbf{A}+\mathbf{A}_p)\cdot (\mathbf{B} - \mathbf{B}_p) dV,
\end{equation}
where $\mathbf{B}_p = \nabla\times\mathbf{A}_p$, but also where $\mathbf{B}_p$ is potential ($\nabla\times\mathbf{B}_p=0$) and is a reference component from $\mathbf{B}$ ($\mathbf{B}_p\cdot\mathbf{n}|_S = \mathbf{B}\cdot\mathbf{n}|_S$).

Thus that we want to find a potential magnetic field $\mathbf{B}_p$ deriving from a scalar magnetic potential $\phi$ ($\mathbf{B}_p = -\nabla\phi$) which we can project on spherical harmonics:
\begin{equation}
\phi(r,\theta,\varphi) = \sum_{\ell,m} \left(A_\ell^mr^\ell + B_\ell^mr^{-(\ell+1)}\right)Y_\ell^m(\theta,\varphi).
\label{eq:phi_ylm}
\end{equation}

Because it is potential $\mathbf{B}_p$ is also poloidal, thus we can introduce a quantity $C_p$ such as:
\begin{equation}
\mathbf{B}_p = \nabla\times\nabla\times(C_p\mathbf{e}_r) \Rightarrow \mathbf{A}_p = \nabla\times(C_p\mathbf{e}_r).
\end{equation}

Thus, to obtain $\mathbf{B}_p$ and $\mathbf{A}_p$, we only need to compute $C_p$, which can be linked to $\phi$ using the following relation:
\begin{equation}
C_p = - \sum_{\ell,m} \left(\frac{A_\ell^m}{\ell+1}r^{\ell+1} - \frac{B_\ell^m}{\ell}r^{-\ell}\right)Y_\ell^m.
\end{equation}

Under the assumption of axisymmetry, we have in the end:
\begin{equation}
\mathbf{A}_p = -\frac{1}{r}\frac{\partial C_p}{\partial\theta}\mathbf{e}_\varphi,
\mathbf{B}_p = -\frac{\partial \phi}{\partial r}\mathbf{e}_r - \frac{1}{r}\frac{\partial\phi}{\partial\theta} \mathbf{e}_\theta,
\end{equation}
\begin{equation}
H_{rel} = \int (\mathbf{A}\cdot\mathbf{B} - \mathbf{A}\cdot\mathbf{B}_p + A_pB_\phi)dV.
\end{equation}

The computation of $A_\ell^m$ and $B_\ell^m$ depends on the boundary conditions of the domain.

\subsubsection{Potential components with perfect conductor and potential field}

For the dynamo, we have a perfect conductor boundary condition at the bottom, which leads to:
\begin{equation}
\partial_r\phi(r=r_b) = 0 \Leftrightarrow \ell r_b^{\ell - 1}A_\ell^m - (\ell+1)r_b^{-(\ell+2)}B_\ell^m = 0.
\end{equation}

At the top we have a potential field boundary condition, which implies:
\begin{equation}
\partial_r\phi (r=r_t) = - B_r(r=r_t) \Leftrightarrow \ell r_t^{\ell - 1}A_\ell^m - (\ell+1)r_t^{-(\ell+2)}B_\ell^m = - G_\ell^m,
\end{equation}
with $B_r(r_t) = \sum_{\ell,m} G_\ell^m Y_\ell^m$.

We combine these two equations to obtain in the end:
\begin{equation}
    A_\ell^m = \frac{G_\ell^m}{\ell r_t^{\ell-1}\left[\left(\frac{r_b}{r_t}\right)^{2\ell+1} - 1\right]},
    B_\ell^m = \frac{r_b^{2\ell+1}G_\ell^m}{(\ell+1) r_t^{\ell-1}\left[\left(\frac{r_b}{r_t}\right)^{2\ell+1} - 1\right]}.
\end{equation}

\subsubsection{Potential components with potential field and outflow boundaries}

For the wind, we have a potential field boundary at the bottom, which leads to:
\begin{equation}
\partial_r\phi (r=r_b) = - B_r(r=r_b) \Leftrightarrow \ell r_b^{\ell - 1}A_\ell^m - (\ell+1)r_b^{-(\ell+2)}B_\ell^m = - G_\ell^m,
\label{eq:pf_wind}
\end{equation}
with $B_r(r_b) = \sum_{\ell,m} G_\ell^m Y_\ell^m$.

At the top we have an outflow boundary condition, which is expressed using a modified outflow boundary condition with $\partial_r(B_rr^2)=0$ instead of $\partial_rB_r=0$, which leads to:
\begin{equation}
2r_tB_r(r=r_t) + r_t^2\partial_rB_r(r=r_t) = 0 \Leftrightarrow 2r_t\partial_r\phi(r=r_t) + r_t^2\partial_r^2\phi(r=r_t) = 0.
\end{equation}

With expression \ref{eq:phi_ylm}, we obtain:
\begin{equation}
 r_t^\ell A_\ell^m + r_t^{-(\ell+1)}B_\ell^m = 0.
\end{equation}

We combine this condition with condition \ref{eq:pf_wind} to obtain in the end:
\begin{equation}
    B_\ell^m = \frac{G_\ell^m r_b^{\ell+2}}{ \left(\frac{r_b}{r_t}\right)^{2\ell+1} \ell + (\ell + 1)}\, ,
    A_\ell^m = - r_t^{-(2\ell+1)} B_\ell^m .
\end{equation}

\subsection{Computation of $\mathbf{A}$}

We have a magnetic field $\mathbf{B}$ which derives from a vector potential $\mathbf{A}$ such that $\mathbf{B} = \nabla \times \mathbf{A}$. Knowing $\mathbf{B}$, we want to recover $\mathbf{A}$ in 2.5D.

To do so, we decompose the magnetic field $\mathbf{B}$ in a poloidal and a toroidal term in spherical coordinates:
\begin{equation}
\mathbf{B} = \nabla\times(C\mathbf{e}_\varphi) + \nabla\times(A\mathbf{e}_r).
\label{eq:b_c_a}
\end{equation}

To obtain $C$, we can use direct integration, given that:
\begin{equation}
    Br = \frac{1}{r\rm{sin}\ \theta}\frac{\partial}{\partial\theta}(\rm{sin}\ \theta C), B_\theta = -\frac{1}{r}\frac{\partial}{\partial r}(rC).
\end{equation}

To obtain $A$, we use two different methods: direct integration for outside the star, projection on spherical harmonics for inside the star.

With direct integration, we have:
\begin{equation}
    B_\varphi = -\frac{1}{r}\partial_\theta A. \\
\end{equation}

We can project $A$ on spherical harmonics, which leads to:
\begin{equation}
A = \sum_{\ell,m} A_\ell^m Y_\ell^m.
\end{equation}

Now if we take the $r$ component of the projection of the current on spherical harmonics, we obtain:
\begin{equation}
\sum_{\ell,m}\frac{\ell(\ell+1)}{r^2}A_\ell^mY_\ell^m = \nabla\times{\bf B}_r.
\end{equation}

We can then link $A$ and $C$ to $\mathbf{A}$ using the following relation:
\begin{equation}
A_r = A, A_\theta = 0, A_\varphi = C.
\end{equation}

\subsection{Balance for helicity fluxes}

To make sure that all the helicity computations were carried out correctly, we have checked the balance described by equation \ref{eq:dhdt}, which we rewrite here under the form:
\begin{equation}
    \frac{\partial H_{rel}}{\partial t} = \Gamma_d + F_H,
\end{equation}
where $\Gamma_d$ is the diffusion term due to the electric currents:
\begin{equation}
    \Gamma_d = -\frac{8\pi \eta}{c} \int \mathbf{j}\cdot\mathbf{B}dV,
\end{equation}
and $F_H$ is the helicity flux:
\begin{equation}
    F_H = 2\int_{\partial V} \left[\mathbf{A}_p\times\left(-\mathbf{u}\times\mathbf{B}+\eta\nabla\times\mathbf{B}\right)\right]\cdot \mathbf{n}dS.
\end{equation}

\begin{figure}
    \centering
    \includegraphics[width=\textwidth]{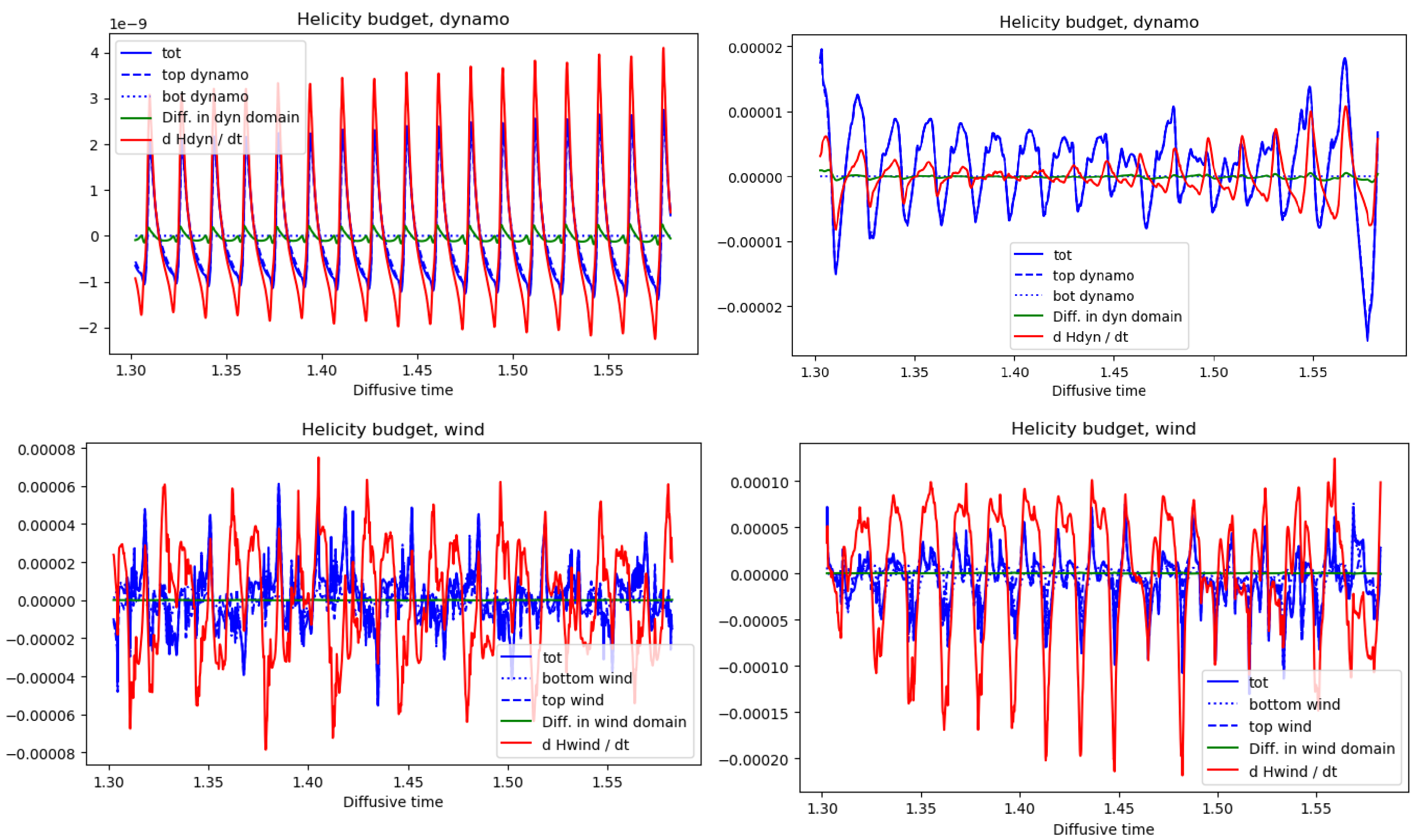}
    \caption{Relative magnetic helicity budgets for case DW1 (on the left) and DW2 (on the right), in code units. Top panel is the budget for the dynamo domain while bottom panel is the budget for the wind domain. For each domain, we plot in red $\partial_tH_{rel}$, in green $\Gamma_d$, in dashed blue the fluxes $F_H$ at the boundaries and in blue the sum $\Gamma_d+F_H$. The balance is qualitatively realized with the blue and red curves following each other in most instances.}
    \label{fig:hel_budget}
\end{figure}

The results are presented in figure \ref{fig:hel_budget}. Case DW1 is on the left, case DW2 on the right. Top panel is the budget for the dynamo domain, while bottom panel is the budget for the wind domain. For each domain, we plot in red $\partial_tH_{rel}$, in green $\Gamma_d$, in dashed blue the fluxes $F_H$ at the boundaries and in blue the sum $\Gamma_d+F_H$. We notice that in almost every cases, the diffusive term $\Gamma_d$ is negligible compared to the other terms. We include it however for more accuracy. The balance is qualitatively realized with the blue and red curves following each other in most instances. What we observe is that there are some minor errors, probably due to the numerical methods used to estimate the relative helicity and to derive the potential vector, and maybe to a lesser extend to small deviations from $\nabla\cdot{\bf B}=0$ that are not corrected fast enough by our divergence cleaning method. The trends are nevertheless in good agreement, giving confidence in our helicity budget analysis.

\bibliography{coupling.bib}

\begin{thebibliography}{}
\expandafter\ifx\csname natexlab\endcsname\relax\def\natexlab#1{#1}\fi
\providecommand{\url}[1]{\href{#1}{#1}}

\bibitem[{Arge \& Pizzo(2000)}]{arge_improvement_2000}
Arge, C.~N., \& Pizzo, V.~J. 2000, Journal of Geophysical Research, 105, 10465.
\newblock \url{https://ui.adsabs.harvard.edu/abs/2000JGR...10510465A/abstract}

\bibitem[{Asai {et~al.}(1998)Asai, Kojima, Tokumaru, Yokobe, Jackson, Hick, \&
  Manoharan}]{asai_heliospheric_1998}
Asai, K., Kojima, M., Tokumaru, M., {et~al.} 1998, Journal of Geophysical
  Research, 103, 1991.
\newblock \url{https://ui.adsabs.harvard.edu/abs/1998JGR...103.1991A/abstract}

\bibitem[{Babcock(1961)}]{babcock_topology_1961}
Babcock, H.~W. 1961, The Astrophysical Journal, 133, 572.
\newblock \url{https://ui.adsabs.harvard.edu/abs/1961ApJ...133..572B/abstract}

\bibitem[{Barnes(1988)}]{barnes_mechanical_1988}
Barnes, D.~C. 1988, The Physics of Fluids, 31, 2214, publisher: American
  Institute of Physics.
\newblock \url{http://aip.scitation.org/doi/10.1063/1.866622}

\bibitem[{Basu \& Antia(2010)}]{basu_characteristics_2010}
Basu, S., \& Antia, H.~M. 2010, The Astrophysical Journal, 717, 488.
\newblock \url{https://ui.adsabs.harvard.edu/abs/2010ApJ...717..488B/abstract}

\bibitem[{Berger(1999)}]{berger_introduction_1999}
Berger, M.~A. 1999, Plasma Phys. Control. Fusion, 41, B167.
\newblock \url{https://iopscience.iop.org/article/10.1088/0741-3335/41/12B/312}

\bibitem[{Browning {et~al.}(2006)Browning, Miesch, Brun, \&
  Toomre}]{browning_dynamo_2006}
Browning, M.~K., Miesch, M.~S., Brun, A.~S., \& Toomre, J. 2006, ApJ, 648,
  L157.
\newblock \url{https://iopscience.iop.org/article/10.1086/507869}

\bibitem[{Brun \& Browning(2017)}]{brun_magnetism_2017}
Brun, A.~S., \& Browning, M.~K. 2017, Living Reviews in Solar Physics, 14, 4.
\newblock \url{https://ui.adsabs.harvard.edu/abs/2017LRSP...14....4B/abstract}

\bibitem[{Charbonneau(2020)}]{charbonneau_dynamo_2020}
Charbonneau, P. 2020, Living Rev Sol Phys, 17, 4.
\newblock \url{http://link.springer.com/10.1007/s41116-020-00025-6}

\bibitem[{Cranmer {et~al.}(2007)Cranmer, van Ballegooijen, \&
  Edgar}]{cranmer_self-consistent_2007}
Cranmer, S.~R., van Ballegooijen, A.~A., \& Edgar, R.~J. 2007, The
  Astrophysical Journal Supplement Series, 171, 520.
\newblock \url{https://ui.adsabs.harvard.edu/abs/2007ApJS..171..520C/abstract}

\bibitem[{Dedner {et~al.}(2002)Dedner, Kemm, Kröner, Munz, Schnitzer, \&
  Wesenberg}]{dedner_hyperbolic_2002}
Dedner, A., Kemm, F., Kröner, D., {et~al.} 2002, Journal of Computational
  Physics, 175, 645.
\newblock \url{https://ui.adsabs.harvard.edu/abs/2002JCoPh.175..645D/abstract}

\bibitem[{DeRosa {et~al.}(2012)DeRosa, Brun, \& Hoeksema}]{derosa_solar_2012}
DeRosa, M.~L., Brun, A.~S., \& Hoeksema, J.~T. 2012, ApJ, 757, 96, arXiv:
  1208.1768.
\newblock \url{http://arxiv.org/abs/1208.1768}

\bibitem[{Dikpati \& Charbonneau(1999)}]{dikpati_babcock-leighton_1999}
Dikpati, M., \& Charbonneau, P. 1999, The Astrophysical Journal, 518, 508.
\newblock \url{https://ui.adsabs.harvard.edu/abs/1999ApJ...518..508D/abstract}

\bibitem[{Einfeldt(1988)}]{einfeldt_godunov-type_1988}
Einfeldt, B. 1988, SIAM Journal on Numerical Analysis, 25, 294.
\newblock \url{https://ui.adsabs.harvard.edu/abs/1988SJNA...25..294E/abstract}

\bibitem[{Finley {et~al.}(2019)Finley, Deshmukh, Matt, Owens, \&
  Wu}]{finley_solar_2019}
Finley, A.~J., Deshmukh, S., Matt, S.~P., Owens, M., \& Wu, C.-J. 2019, The
  Astrophysical Journal, 883, 67.
\newblock \url{http://adsabs.harvard.edu/abs/2019ApJ...883...67F}

\bibitem[{Gary(2001)}]{gary_plasma_2001}
Gary, G.~A. 2001, Solar Physics, 203, 71.
\newblock \url{https://ui.adsabs.harvard.edu/abs/2001SoPh..203...71G/abstract}

\bibitem[{Hale(1908)}]{hale_zeeman_1908}
Hale, G.~E. 1908, Publications of the Astronomical Society of the Pacific, 20,
  287.
\newblock \url{https://ui.adsabs.harvard.edu/abs/1908PASP...20..287H/abstract}

\bibitem[{Hazra {et~al.}(2021)Hazra, Réville, Perri, Strugarek, Brun, \&
  Buchlin}]{hazra_modeling_2021}
Hazra, S., Réville, V., Perri, B., {et~al.} 2021, arXiv:2101.11511 [astro-ph,
  physics:physics], arXiv: 2101.11511.
\newblock \url{http://arxiv.org/abs/2101.11511}

\bibitem[{Hewish {et~al.}(1964)Hewish, Scott, \&
  Wills}]{hewish_interplanetary_1964}
Hewish, A., Scott, P.~F., \& Wills, D. 1964, Nature, 203, 1214.
\newblock \url{https://ui.adsabs.harvard.edu/abs/1964Natur.203.1214H/abstract}

\bibitem[{Hollweg \& Isenberg(2002)}]{hollweg_generation_2002}
Hollweg, J.~V., \& Isenberg, P.~A. 2002, Journal of Geophysical Research (Space
  Physics), 107, 1147.
\newblock \url{https://ui.adsabs.harvard.edu/abs/2002JGRA..107.1147H/abstract}

\bibitem[{Jouve \& Brun(2007)}]{jouve_role_2007}
Jouve, L., \& Brun, A.~S. 2007, Astronomy and Astrophysics, 474, 239.
\newblock
  \url{https://ui.adsabs.harvard.edu/abs/2007A%26A...474..239J/abstract}

\bibitem[{Jouve {et~al.}(2008)Jouve, Brun, Arlt, Brandenburg, Dikpati, Bonanno,
  Käpylä, Moss, Rempel, Gilman, Korpi, \& Kosovichev}]{jouve_solar_2008}
Jouve, L., Brun, A.~S., Arlt, R., {et~al.} 2008, Astronomy and Astrophysics,
  483, 949.
\newblock
  \url{https://ui.adsabs.harvard.edu/abs/2008A%26A...483..949J/abstract}

\bibitem[{Karak \& Miesch(2017)}]{karak_solar_2017}
Karak, B.~B., \& Miesch, M. 2017, ApJ, 847, 69.
\newblock \url{https://iopscience.iop.org/article/10.3847/1538-4357/aa8636}

\bibitem[{Keppens \& Goedbloed(1999)}]{keppens_numerical_1999}
Keppens, R., \& Goedbloed, J.~P. 1999, Astronomy and Astrophysics, 343, 251.
\newblock
  \url{https://ui.adsabs.harvard.edu/abs/1999A%26A...343..251K/abstract}

\bibitem[{Krause \& Raedler(1980)}]{krause_mean-field_1980}
Krause, F., \& Raedler, K.-H. 1980, Oxford.
\newblock \url{https://ui.adsabs.harvard.edu/abs/1980opp..bookR....K/abstract}

\bibitem[{Kumar {et~al.}(2019)Kumar, Jouve, \& Nandy}]{kumar_3d_2019}
Kumar, R., Jouve, L., \& Nandy, D. 2019, A\&A, 623, A54, arXiv: 1901.04251.
\newblock \url{http://arxiv.org/abs/1901.04251}

\bibitem[{Leighton(1969)}]{leighton_magneto-kinematic_1969}
Leighton, R.~B. 1969, The Astrophysical Journal, 156, 1.
\newblock \url{https://ui.adsabs.harvard.edu/abs/1969ApJ...156....1L/abstract}

\bibitem[{Lionello {et~al.}(2001)Lionello, Linker, \&
  Mikić}]{lionello_including_2001}
Lionello, R., Linker, J.~A., \& Mikić, Z. 2001, The Astrophysical Journal,
  546, 542.
\newblock \url{https://ui.adsabs.harvard.edu/abs/2001ApJ...546..542L/abstract}

\bibitem[{Low(2013)}]{low_magnetic_2013}
Low, B.~C. 2013, in Magnetic {Helicity} in {Space} and {Laboratory} {Plasmas}
  (American Geophysical Union (AGU)), 25--32, \_eprint:
  https://onlinelibrary.wiley.com/doi/pdf/10.1029/GM111p0025.
\newblock
  \url{http://agupubs.onlinelibrary.wiley.com/doi/abs/10.1029/GM111p0025}

\bibitem[{Luhmann {et~al.}(2002)Luhmann, Li, Arge, Gazis, \&
  Ulrich}]{luhmann_solar_2002}
Luhmann, J.~G., Li, Y., Arge, C.~N., Gazis, P.~R., \& Ulrich, R. 2002, Journal
  of Geophysical Research (Space Physics), 107, 1154.
\newblock \url{https://ui.adsabs.harvard.edu/abs/2002JGRA..107.1154L/abstract}

\bibitem[{Matt \& Pudritz(2008)}]{matt_accretion-powered_2008}
Matt, S., \& Pudritz, R.~E. 2008, The Astrophysical Journal, 678, 1109.
\newblock \url{https://ui.adsabs.harvard.edu/abs/2008ApJ...678.1109M/abstract}

\bibitem[{McComas {et~al.}(2008)McComas, Ebert, Elliott, Goldstein, Gosling,
  Schwadron, \& Skoug}]{mccomas_weaker_2008}
McComas, D.~J., Ebert, R.~W., Elliott, H.~A., {et~al.} 2008, Geophysical
  Research Letters, 35, L18103.
\newblock \url{https://ui.adsabs.harvard.edu/abs/2008GeoRL..3518103M/abstract}

\bibitem[{McComas {et~al.}(2003)McComas, Elliott, Schwadron, Gosling, Skoug, \&
  Goldstein}]{mccomas_three-dimensional_2003}
McComas, D.~J., Elliott, H.~A., Schwadron, N.~A., {et~al.} 2003, Geophysical
  Research Letters, 30, 1517.
\newblock \url{https://ui.adsabs.harvard.edu/abs/2003GeoRL..30.1517M/abstract}

\bibitem[{McFadden {et~al.}(1991)McFadden, Merrill, McElhinny, \&
  Lee}]{mcfadden_reversals_1991}
McFadden, P.~L., Merrill, R.~T., McElhinny, M.~W., \& Lee, S. 1991, Journal of
  Geophysical Research: Solid Earth, 96, 3923, \_eprint:
  https://onlinelibrary.wiley.com/doi/pdf/10.1029/90JB02275.
\newblock
  \url{http://agupubs.onlinelibrary.wiley.com/doi/abs/10.1029/90JB02275}

\bibitem[{Merkin {et~al.}(2016)Merkin, Lyon, Lario, Arge, \&
  Henney}]{merkin_time-dependent_2016}
Merkin, V.~G., Lyon, J.~G., Lario, D., Arge, C.~N., \& Henney, C.~J. 2016,
  Journal of Geophysical Research (Space Physics), 121, 2866.
\newblock \url{https://ui.adsabs.harvard.edu/abs/2016JGRA..121.2866M/abstract}

\bibitem[{Mestel(1968)}]{mestel_magnetic_1968}
Mestel, L. 1968, Monthly Notices of the Royal Astronomical Society, 140, 177.
\newblock
  \url{https://academic.oup.com/mnras/article-lookup/doi/10.1093/mnras/140.2.177}

\bibitem[{Meyer-Vernet(2007)}]{meyer-vernet_basics_2007}
Meyer-Vernet, N. 2007, Basics of the Solar Wind.
\newblock \url{https://ui.adsabs.harvard.edu/abs/2007bsw..book.....M/abstract}

\bibitem[{Miesch(2005)}]{miesch_large-scale_2005}
Miesch, M.~S. 2005, Living Reviews in Solar Physics, 2, 1.
\newblock \url{https://ui.adsabs.harvard.edu/abs/2005LRSP....2....1M/abstract}

\bibitem[{Mignone {et~al.}(2007)Mignone, Bodo, Massaglia, Matsakos, Tesileanu,
  Zanni, \& Ferrari}]{mignone_pluto:_2007}
Mignone, A., Bodo, G., Massaglia, S., {et~al.} 2007, The Astrophysical Journal
  Supplement Series, 170, 228.
\newblock \url{https://ui.adsabs.harvard.edu/abs/2007ApJS..170..228M/abstract}

\bibitem[{Moffatt(1978)}]{moffatt_magnetic_1978}
Moffatt, H.~K. 1978, Cambridge Monographs on Mechanics and Applied Mathematics.
\newblock \url{https://ui.adsabs.harvard.edu/abs/1978mfge.book.....M/abstract}

\bibitem[{Neugebauer \& Snyder(1962)}]{neugebauer_solar_1962}
Neugebauer, M., \& Snyder, C.~W. 1962, Science, 138, 1095.
\newblock \url{https://ui.adsabs.harvard.edu/abs/1962Sci...138.1095N/abstract}

\bibitem[{Ossendrijver(2003)}]{ossendrijver_solar_2003}
Ossendrijver, M. 2003, Astronomy and Astrophysics Review, 11, 287.
\newblock
  \url{https://ui.adsabs.harvard.edu/abs/2003A%26ARv..11..287O/abstract}

\bibitem[{Owens {et~al.}(2017)Owens, Lockwood, \& Riley}]{owens_global_2017}
Owens, M.~J., Lockwood, M., \& Riley, P. 2017, Scientific Reports, 7, 41548.
\newblock \url{https://ui.adsabs.harvard.edu/abs/2017NatSR...741548O/abstract}

\bibitem[{Parker(1958)}]{parker_dynamics_1958}
Parker, E.~N. 1958, The Astrophysical Journal, 128, 664.
\newblock \url{https://ui.adsabs.harvard.edu/abs/1958ApJ...128..664P/abstract}

\bibitem[{Parker(1988)}]{parker_nanoflares_1988}
---. 1988, The Astrophysical Journal, 330, 474.
\newblock \url{https://ui.adsabs.harvard.edu/abs/1988ApJ...330..474P/abstract}

\bibitem[{Parker(1993)}]{parker_solar_1993}
---. 1993, The Astrophysical Journal, 408, 707.
\newblock \url{https://ui.adsabs.harvard.edu/abs/1993ApJ...408..707P/abstract}

\bibitem[{Perri {et~al.}(2018)Perri, Brun, Réville, \&
  Strugarek}]{perri_simulations_2018}
Perri, B., Brun, A.~S., Réville, V., \& Strugarek, A. 2018, Journal of Plasma
  Physics, 84, doi:10.1017/S0022377818000880.
\newblock
  \url{https://www.cambridge.org/core/journals/journal-of-plasma-physics/article/simulations-of-solar-wind-variations-during-an-11year-cycle-and-the-influence-of-northsouth-asymmetry/D1E8A3463A3F9F113BC21E63EAEA28E0#}

\bibitem[{Pinto {et~al.}(2011)Pinto, Brun, Jouve, \&
  Grappin}]{pinto_coupling_2011}
Pinto, R.~F., Brun, A.~S., Jouve, L., \& Grappin, R. 2011, The Astrophysical
  Journal, 737, 72.
\newblock \url{https://ui.adsabs.harvard.edu/abs/2011ApJ...737...72P/abstract}

\bibitem[{Pinto \& Rouillard(2017)}]{pinto_multiple_2017}
Pinto, R.~F., \& Rouillard, A.~P. 2017, The Astrophysical Journal, 838, 89.
\newblock \url{https://ui.adsabs.harvard.edu/abs/2017ApJ...838...89P/abstract}

\bibitem[{Pouquet {et~al.}(1976)Pouquet, Frisch, \&
  Leorat}]{pouquet_strong_1976}
Pouquet, A., Frisch, U., \& Leorat, J. 1976, Journal of Fluid Mechanics, 77,
  321.
\newblock \url{https://ui.adsabs.harvard.edu/abs/1976JFM....77..321P/abstract}

\bibitem[{Rieutord \& Rincon(2010)}]{rieutord_suns_2010}
Rieutord, M., \& Rincon, F. 2010, Living Reviews in Solar Physics, 7, 2.
\newblock \url{https://ui.adsabs.harvard.edu/abs/2010LRSP....7....2R/abstract}

\bibitem[{Riley {et~al.}(2015)Riley, Lionello, Linker, Cliver, Balogh, Beer,
  Charbonneau, Crooker, DeRosa, Lockwood, Owens, McCracken, Usoskin, \&
  Koutchmy}]{riley_inferring_2015}
Riley, P., Lionello, R., Linker, J.~A., {et~al.} 2015, The Astrophysical
  Journal, 802, 105.
\newblock \url{https://ui.adsabs.harvard.edu/abs/2015ApJ...802..105R/abstract}

\bibitem[{Roberts(1972)}]{roberts_kinematic_1972}
Roberts, P.~H. 1972, Philosophical Transactions of the Royal Society of London
  Series A, 272, 663.
\newblock \url{https://ui.adsabs.harvard.edu/abs/1972RSPTA.272..663R/abstract}

\bibitem[{Réville \& Brun(2017)}]{reville_global_2017}
Réville, V., \& Brun, A.~S. 2017, The Astrophysical Journal, 850, 45.
\newblock \url{https://ui.adsabs.harvard.edu/abs/2017ApJ...850...45R/abstract}

\bibitem[{Réville {et~al.}(2015)Réville, Brun, Strugarek, Matt, Bouvier,
  Folsom, \& Petit}]{reville_solar_2015}
Réville, V., Brun, A.~S., Strugarek, A., {et~al.} 2015, The Astrophysical
  Journal, 814, 99.
\newblock \url{https://ui.adsabs.harvard.edu/abs/2015ApJ...814...99R/abstract}

\bibitem[{Réville {et~al.}(2018)Réville, Tenerani, \&
  Velli}]{reville_parametric_2018}
Réville, V., Tenerani, A., \& Velli, M. 2018, The Astrophysical Journal, 866,
  38.
\newblock \url{https://ui.adsabs.harvard.edu/abs/2018ApJ...866...38R/abstract}

\bibitem[{Réville {et~al.}(2020)Réville, Velli, Panasenco, Tenerani, Shi,
  Badman, Bale, Kasper, Stevens, Korreck, Bonnell, Case, de~Wit, Goetz, Harvey,
  Larson, Livi, Malaspina, MacDowall, Pulupa, \&
  Whittlesey}]{reville_role_2020}
Réville, V., Velli, M., Panasenco, O., {et~al.} 2020, ApJS, 246, 24.
\newblock \url{https://iopscience.iop.org/article/10.3847/1538-4365/ab4fef}

\bibitem[{Sakurai(1985)}]{sakurai_magnetic_1985}
Sakurai, T. 1985, Astronomy and Astrophysics, 152, 121.
\newblock
  \url{https://ui.adsabs.harvard.edu/abs/1985A%26A...152..121S/abstract}

\bibitem[{Schatzman(1962)}]{schatzman_theory_1962}
Schatzman, E. 1962, Annales d'Astrophysique, 25, 18.
\newblock \url{https://ui.adsabs.harvard.edu/abs/1962AnAp...25...18S/abstract}

\bibitem[{Schou \& Bogart(1998)}]{schou_flows_1998}
Schou, J., \& Bogart, R.~S. 1998, The Astrophysical Journal, 504, L131.
\newblock \url{https://ui.adsabs.harvard.edu/abs/1998ApJ...504L.131S/abstract}

\bibitem[{Schrijver(2005)}]{schrijver_magnetic_2005}
Schrijver, C.~J. 2005, Solar Wind 11/SOHO 16, Connecting Sun and Heliosphere,
  592, 213.
\newblock \url{https://ui.adsabs.harvard.edu/abs/2005ESASP.592..213S/abstract}

\bibitem[{Shoda {et~al.}(2020)Shoda, Suzuki, Matt, Cranmer, Vidotto, Strugarek,
  See, Réville, Finley, \& Brun}]{shoda_alfven-wave_2020}
Shoda, M., Suzuki, T.~K., Matt, S.~P., {et~al.} 2020, ApJ, 896, 123, arXiv:
  2005.09817.
\newblock \url{http://arxiv.org/abs/2005.09817}

\bibitem[{Skartlien {et~al.}(2000)Skartlien, Stein, \&
  Nordlund}]{skartlien_excitation_2000}
Skartlien, R., Stein, R.~F., \& Nordlund, A. 2000, The Astrophysical Journal,
  541, 468.
\newblock \url{https://ui.adsabs.harvard.edu/abs/2000ApJ...541..468S/abstract}

\bibitem[{Sokół {et~al.}(2015)Sokół, Swaczyna, Bzowski, \&
  Tokumaru}]{sokol_reconstruction_2015}
Sokół, J.~M., Swaczyna, P., Bzowski, M., \& Tokumaru, M. 2015, Solar Physics,
  290, 2589.
\newblock \url{https://ui.adsabs.harvard.edu/abs/2015SoPh..290.2589S/abstract}

\bibitem[{Spiegel \& Zahn(1992)}]{spiegel_solar_1992}
Spiegel, E.~A., \& Zahn, J.-P. 1992, Astronomy and Astrophysics, 265, 106.
\newblock
  \url{https://ui.adsabs.harvard.edu/abs/1992A%26A...265..106S/abstract}

\bibitem[{Stein \& Nordlund(2006)}]{stein_solar_2006}
Stein, R.~F., \& Nordlund, A. 2006, The Astrophysical Journal, 642, 1246.
\newblock \url{https://ui.adsabs.harvard.edu/abs/2006ApJ...642.1246S/abstract}

\bibitem[{Suzuki {et~al.}(2013)Suzuki, Imada, Kataoka, Kato, Matsumoto,
  Miyahara, \& Tsuneta}]{suzuki_saturation_2013}
Suzuki, T.~K., Imada, S., Kataoka, R., {et~al.} 2013, Publications of the
  Astronomical Society of Japan, 65, 98.
\newblock \url{https://ui.adsabs.harvard.edu/abs/2013PASJ...65...98S/abstract}

\bibitem[{Svalgaard \& Kamide(2013)}]{svalgaard_asymmetric_2013}
Svalgaard, L., \& Kamide, Y. 2013, The Astrophysical Journal, 763, 23.
\newblock \url{https://ui.adsabs.harvard.edu/abs/2013ApJ...763...23S/abstract}

\bibitem[{Tavakol {et~al.}(1995)Tavakol, Tworkowski, Brandenburg, Moss, \&
  Tuominen}]{tavakol_structural_1995}
Tavakol, R., Tworkowski, A.~S., Brandenburg, A., Moss, D., \& Tuominen, I.
  1995, Astronomy and Astrophysics, 296, 269.
\newblock
  \url{https://ui.adsabs.harvard.edu/abs/1995A%26A...296..269T/abstract}

\bibitem[{Thompson {et~al.}(2003)Thompson, Christensen-Dalsgaard, Miesch, \&
  Toomre}]{thompson_internal_2003}
Thompson, M.~J., Christensen-Dalsgaard, J., Miesch, M.~S., \& Toomre, J. 2003,
  Annual Review of Astronomy and Astrophysics, 41, 599.
\newblock
  \url{https://ui.adsabs.harvard.edu/abs/2003ARA%26A..41..599T/abstract}

\bibitem[{Tobias(2019)}]{tobias_turbulent_2019}
Tobias, S. 2019, arXiv e-prints, arXiv:1907.03685.
\newblock \url{https://ui.adsabs.harvard.edu/abs/2019arXiv190703685T/abstract}

\bibitem[{Tokumaru {et~al.}(2010)Tokumaru, Kojima, \&
  Fujiki}]{tokumaru_solar_2010}
Tokumaru, M., Kojima, M., \& Fujiki, K. 2010, Journal of Geophysical Research
  (Space Physics), 115, A04102.
\newblock \url{https://ui.adsabs.harvard.edu/abs/2010JGRA..115.4102T/abstract}

\bibitem[{Tóth {et~al.}(2012)Tóth, van~der Holst, Sokolov, De~Zeeuw, Gombosi,
  Fang, Manchester, Meng, Najib, Powell, Stout, Glocer, Ma, \&
  Opher}]{toth_adaptive_2012}
Tóth, G., van~der Holst, B., Sokolov, I.~V., {et~al.} 2012, Journal of
  Computational Physics, 231, 870.
\newblock \url{https://ui.adsabs.harvard.edu/abs/2012JCoPh.231..870T/abstract}

\bibitem[{Usmanov {et~al.}(2000)Usmanov, Goldstein, Besser, \&
  Fritzer}]{usmanov_global_2000}
Usmanov, A.~V., Goldstein, M.~L., Besser, B.~P., \& Fritzer, J.~M. 2000,
  Journal of Geophysical Research, 105, 12675.
\newblock \url{https://ui.adsabs.harvard.edu/abs/2000JGR...10512675U/abstract}

\bibitem[{Usmanov {et~al.}(2014)Usmanov, Goldstein, \&
  Matthaeus}]{usmanov_three-fluid_2014}
Usmanov, A.~V., Goldstein, M.~L., \& Matthaeus, W.~H. 2014, AGU Fall Meeting
  Abstracts, 2014, SH53C.
\newblock \url{https://ui.adsabs.harvard.edu/abs/2014AGUFMSH53C..04U/abstract}

\bibitem[{Vernazza {et~al.}(1981)Vernazza, Avrett, \&
  Loeser}]{vernazza_structure_1981}
Vernazza, J.~E., Avrett, E.~H., \& Loeser, R. 1981, The Astrophysical Journal
  Supplement Series, 45, 635.
\newblock \url{https://ui.adsabs.harvard.edu/abs/1981ApJS...45..635V/abstract}

\bibitem[{von Rekowski \& Brandenburg(2006)}]{von_rekowski_stellar_2006}
von Rekowski, B., \& Brandenburg, A. 2006, Astronomische Nachrichten, 327, 53.
\newblock \url{https://ui.adsabs.harvard.edu/abs/2006AN....327...53V/abstract}

\bibitem[{Vögler {et~al.}(2005)Vögler, Shelyag, Schüssler, Cattaneo, Emonet,
  \& Linde}]{vogler_simulations_2005}
Vögler, A., Shelyag, S., Schüssler, M., {et~al.} 2005, Astronomy and
  Astrophysics, 429, 335.
\newblock
  \url{https://ui.adsabs.harvard.edu/abs/2005A%26A...429..335V/abstract}

\bibitem[{Wang \& Sheeley(1990)}]{wang_magnetic_1990}
Wang, Y.-M., \& Sheeley, N.~R. 1990, The Astrophysical Journal, 365, 372.
\newblock \url{https://ui.adsabs.harvard.edu/abs/1990ApJ...365..372W/abstract}

\bibitem[{Wang {et~al.}(2007)Wang, Sheeley, \& Rich}]{wang_coronal_2007}
Wang, Y.-M., Sheeley, N.~R., \& Rich, N.~B. 2007, The Astrophysical Journal,
  658, 1340.
\newblock \url{https://ui.adsabs.harvard.edu/abs/2007ApJ...658.1340W/abstract}

\bibitem[{Warnecke {et~al.}(2016)Warnecke, Käpylä, Käpylä, \&
  Brandenburg}]{warnecke_influence_2016}
Warnecke, J., Käpylä, P.~J., Käpylä, M.~J., \& Brandenburg, A. 2016,
  Astronomy and Astrophysics, 596, A115.
\newblock
  \url{https://ui.adsabs.harvard.edu/abs/2016A%26A...596A.115W/abstract}

\bibitem[{Weber \& Davis(1967)}]{weber_angular_1967}
Weber, E.~J., \& Davis, L. 1967, The Astrophysical Journal, 148, 217.
\newblock \url{https://ui.adsabs.harvard.edu/abs/1967ApJ...148..217W/abstract}

\bibitem[{Wedemeyer-Böhm {et~al.}(2009)Wedemeyer-Böhm, Lagg, \&
  Nordlund}]{wedemeyer-bohm_coupling_2009}
Wedemeyer-Böhm, S., Lagg, A., \& Nordlund, A. 2009, Space Science Reviews,
  144, 317.
\newblock \url{https://ui.adsabs.harvard.edu/abs/2009SSRv..144..317W/abstract}

\bibitem[{Yoshimura(1975)}]{yoshimura_model_1975}
Yoshimura, H. 1975, The Astrophysical Journal Supplement Series, 29, 467.
\newblock \url{http://adsabs.harvard.edu/abs/1975ApJS...29..467Y}

\end{thebibliography}

\end{document}